\documentstyle[12pt,epsfig,rotate]{article}
\begin{document}

\begin{titlepage}

\vspace*{-1.5cm}

\begin{flushright}
CERN-TH/99-242\\
hep-ph/9908340
\end{flushright}
 
\vspace*{0.5cm}

\boldmath 
\begin{center}
\Large\bf Recent Theoretical Developments in \\
CP Violation in the $B$ System
\end{center}
\unboldmath 

\vspace{0.5cm}
 
\begin{center}
Robert Fleischer~\footnote{Robert.Fleischer@cern.ch}\\
{\sl Theory Division, CERN, CH-1211 Geneva 23, Switzerland}
\end{center}
 
\vspace{0.5cm}

\begin{center}
{\bf Abstract}\\[0.3cm]
\parbox{11cm}{
After a brief review of the present status of the standard methods to
extract CKM phases from CP-violating effects in non-leptonic $B$-decays,
an overview of recent theoretical developments in this field is given, 
including extractions of $\gamma$ from $B\to\pi K$ and 
$B_{s(d)}\to J/\psi\, K_{\rm S}$ decays, a simultaneous determination
of $\beta$ and $\gamma$, which is provided by the modes $B_d\to \pi^+\pi^-$ 
and $B_s\to K^+K^-$, and extractions of CKM phases from angular 
distributions of certain $B_{d,s}$ decays, such as $B_d\to J/\psi\,\rho^0$
and $B_s\to J/\psi\,\phi$.}
\end{center}
 
\vspace{0.5cm}
 
\begin{center}
{\small{\sl Invited talk given at the\\
6th International Conference on $B$-Physics at Hadron Machines,\\
Bled, Slovenia, 21--25 June 1999\\
To appear in the Proceedings}}
\end{center}

\vfil
\noindent
CERN-TH/99-242\\
August 1999
 
\end{titlepage}
 
\thispagestyle{empty}
\vbox{}
\newpage
 
\setcounter{page}{1}
 

\section{Setting the Stage}

CP violation is one of the central and fundamental phenomena in modern 
particle physics, providing a very fertile testing ground for the 
Standard Model. In this respect, the $B$-meson system plays an outstanding
role, which is also reflected in the tremendous experimental effort put
in the preparations to explore $B$ physics. The BaBar (SLAC) and BELLE (KEK) 
detectors have already seen their first events -- which manifests the 
beginning of the $B$-factory era in particle physics -- and CLEO-III 
(Cornell), HERA-B (DESY) and CDF-II (Fermilab) will start taking data in 
the near future. Although the physics potential of these experiments is 
very promising, it may well be that the ``definite'' answer in the search 
for new physics will be left for second-generation $B$-physics experiments 
at hadron machines, such as LHCb (CERN) or BTeV (Fermilab), which offer, 
among other things, very exciting ways of using $B_s$ decays.

\begin{figure}
\begin{tabular}{lr}
   \epsfysize=3.9cm
   \epsffile{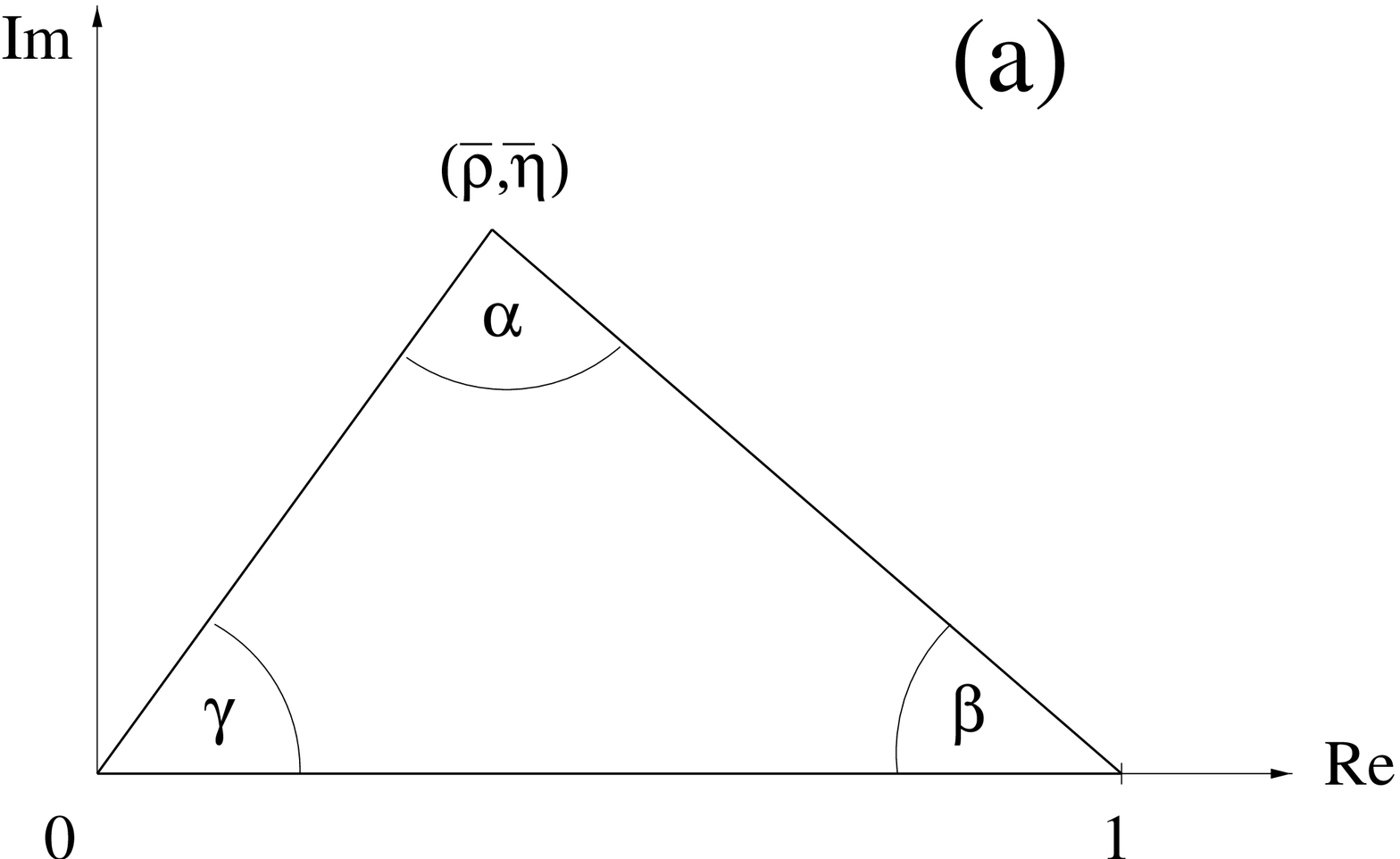}
&
   \epsfysize=3.9cm
   \epsffile{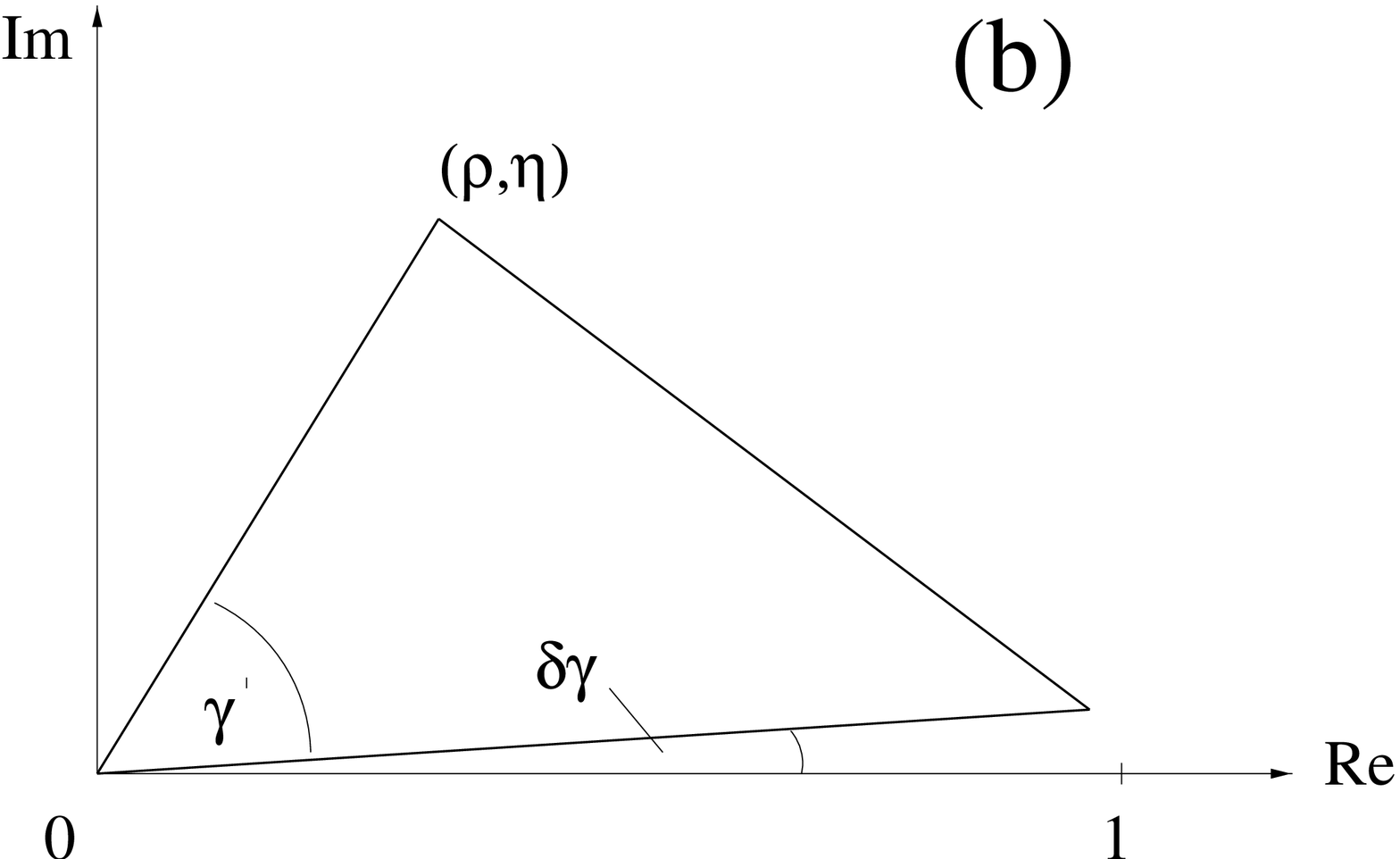}
\end{tabular}
\caption[]{The two non-squashed unitarity triangles of the CKM matrix: 
(a) and (b) correspond to the orthogonality relations (\ref{UT1}) and 
(\ref{UT2}), respectively.}
\label{fig:UT}
\end{figure}

Within the framework of the Standard Model, CP violation is closely related 
to the Cabibbo--Kobayashi--Maskawa (CKM) matrix \cite{ckm}, connecting 
the electroweak eigenstates of the down, strange and bottom quarks with their 
mass eigenstates. As far as CP violation is concerned, the central feature 
is that -- in addition to three generalized Cabibbo-type angles -- also 
a {\it complex phase} is needed in the three-generation case to parametrize 
the CKM matrix. This complex phase is the origin of CP violation within 
the Standard Model. Concerning tests of the CKM picture of CP violation,
the central targets are the {\it unitarity triangles} of the CKM matrix. 
The unitarity of the CKM matrix, which is described by
\begin{equation}
\hat V_{\mbox{{\scriptsize CKM}}}^{\,\,\dagger}\cdot\hat 
V_{\mbox{{\scriptsize CKM}}}=
\hat 1=\hat V_{\mbox{{\scriptsize CKM}}}\cdot\hat V_{\mbox{{\scriptsize 
CKM}}}^{\,\,\dagger},
\end{equation}
leads to a set of 12 equations, consisting of 6 normalization relations 
and 6 orthogonality relations. The latter can be represented as 6 triangles
in the complex plane, all having the same area \cite{AKL}. However, in only 
two of them, all three sides are of comparable magnitude 
${\cal O}(\lambda^3)$, while in the remaining ones, one side is suppressed 
relative to the others by ${\cal O}(\lambda^2)$ or ${\cal O}(\lambda^4)$,
where $\lambda\equiv|V_{us}|=0.22$ denotes the Wolfenstein parameter 
\cite{wolf}. The orthogonality relations describing the non-squashed 
triangles are given as follows:
\begin{eqnarray}
V_{ud}\,V_{ub}^\ast+V_{cd}\,V_{cb}^\ast+V_{td}\,V_{tb}^\ast&=&0\label{UT1}\\
V_{ud}^\ast\, V_{td}+V_{us}^\ast\, V_{ts}+V_{ub}^\ast\, V_{tb}&=&0.\label{UT2}
\end{eqnarray}
The two non-squashed triangles agree at leading order in the Wolfenstein 
expansion (${\cal O}(\lambda^3)$), so that we actually have to 
deal with a single triangle at this order, which is usually referred to as 
``the'' unitarity triangle of the CKM matrix \cite{ut}. However, in the era 
of second-generation experiments, starting around 2005, we will have
to take into account the next-to-leading order terms of the Wolfenstein 
expansion, and will have to distinguish between the unitarity triangles 
described by (\ref{UT1}) and (\ref{UT2}), which are illustrated in 
Fig.\ \ref{fig:UT}. Here, $\overline{\rho}$ and $\overline{\eta}$ are 
related to the Wolfenstein parameters $\rho$ and $\eta$ through \cite{BLO}
\begin{equation}
\overline{\rho}\equiv\left(1-\lambda^2/2\right)\rho,\quad
\overline{\eta}\equiv\left(1-\lambda^2/2\right)\eta,
\end{equation}
and the angle $\delta\gamma=\lambda^2\eta$ in Fig.\ \ref{fig:UT}\,(b) 
measures the CP-violating weak $B^0_s$--$\overline{B^0_s}$ mixing phase, 
as we will see in Subsection \ref{sec:CP-neut}.

The outline of this paper is as follows: in Section~\ref{sec:Stan-Meth},
the standard methods to extract CKM phases from CP-violating effects in 
non-leptonic $B$ decays are reviewed briefly in the light of recent 
theoretical and experimental results. In Section~\ref{sec:New-Strat}, we 
then focus on new theoretical developments in this field, including 
extractions of $\gamma$ from $B\to\pi K$ and $B_{s(d)}\to J/\psi\, K_{\rm S}$ 
decays, a simultaneous determination of $\beta$ and $\gamma$, which is 
provided by the modes $B_d\to \pi^+\pi^-$ and $B_s\to K^+K^-$, and 
extractions of CKM phases and hadronic parameters from angular distributions 
of certain $B_{d,s}$ decays, such as $B_d\to J/\psi\,\rho^0$ and 
$B_s\to J/\psi\,\phi$. Finally, in Section~\ref{sec:concl} we summarize 
the conclusions and give a brief outlook.

\section{A Brief Look at the Standard Methods to Extract 
CKM Phases}\label{sec:Stan-Meth}

In order to determine the angles of the unitarity triangles shown in
Fig.~\ref{fig:UT} and to test the Standard-Model description of CP 
violation, the major role is played by non-leptonic $B$ decays, which 
can be divided into three decay classes: decays receiving both ``tree'' and 
``penguin'' contributions, pure ``tree'' decays, and pure ``penguin'' 
decays. There are two types of penguin topologies: gluonic (QCD) and 
electroweak (EW) penguins, which are related to strong and electroweak 
interactions, respectively. Because of the large top-quark mass, also EW 
penguins play an important role in several processes \cite{rev}. An 
outstanding tool to extract CKM phases is provided by CP-violating effects 
in non-leptonic decays of neutral $B$-mesons.

\boldmath
\subsection{CP Violation in Neutral $B$ Decays}\label{sec:CP-neut}
\unboldmath

A particularly simple and interesting situation arises if we restrict 
ourselves to decays of neutral $B_q$-mesons ($q\in\{d,s\}$) into CP 
self-conjugate final states $|f\rangle$, satisfying the relation 
$({\cal CP})|f\rangle=\pm\,|f\rangle$. In this case, the corresponding 
time-dependent CP asymmetry can be expressed as
\begin{eqnarray}
\lefteqn{a_{\rm CP}(t)\equiv\frac{\Gamma(B^0_q(t)\to f)-
\Gamma(\overline{B^0_q}(t)\to f)}{\Gamma(B^0_q(t)\to f)+
\Gamma(\overline{B^0_q}(t)\to f)}=}\nonumber\\
&&2\,e^{-\Gamma_q t}\left[\frac{{\cal A}_{\rm CP}^{\rm dir}(B_q\to f)
\cos(\Delta M_q t)+{\cal A}_{\rm CP}^{\rm mix}(B_q\to f)\sin(\Delta M_q t)}{
e^{-\Gamma_{\rm H}^{(q)}t}+e^{-\Gamma_{\rm L}^{(q)}t}+
{\cal A}_{\rm \Delta\Gamma}(B_q\to f)\left(e^{-\Gamma_{\rm H}^{(q)}t}-
e^{-\Gamma_{\rm L}^{(q)}t}\right)} \right],\label{ee6}
\end{eqnarray}
where $\Delta M_q\equiv M_{\rm H}^{(q)}-M_{\rm L}^{(q)}$ denotes the
mass difference between the $B_q$ mass eigenstates, and 
$\Gamma_{\rm H,L}^{(q)}$ are the corresponding decay widths, with 
$\Gamma_q\equiv\left(\Gamma_{\rm H}^{(q)}+\Gamma_{\rm L}^{(q)}\right)/2$. 
In Eq.\ (\ref{ee6}), we have separated the ``direct'' from the 
``mixing-induced'' CP-violating contributions, which are described by
\begin{equation}\label{ee7}
{\cal A}^{\mbox{{\scriptsize dir}}}_{\mbox{{\scriptsize CP}}}(B_q\to f)\equiv
\frac{1-\bigl|\xi_f^{(q)}\bigr|^2}{1+\bigl|\xi_f^{(q)}\bigr|^2}\quad
\mbox{and}\quad
{\cal A}^{\mbox{{\scriptsize mix--ind}}}_{\mbox{{\scriptsize
CP}}}(B_q\to f)\equiv\frac{2\,\mbox{Im}\,\xi^{(q)}_f}{1+\bigl|\xi^{(q)}_f
\bigr|^2}\,,
\end{equation}
respectively. Here direct CP violation refers to CP-violating effects
arising directly in the corresponding decay amplitudes, whereas 
mixing-induced CP violation is due to interference effects between 
$B_q^0$--$\overline{B_q^0}$ mixing and decay processes. Whereas the
width difference $\Delta\Gamma_q\equiv\Gamma_{\rm H}^{(q)}-
\Gamma_{\rm L}^{(q)}$ is negligibly small in the $B_d$ system, it may be 
sizeable in the $B_s$ system \cite{dun,DGamma-cal}, thereby providing the 
observable 
\begin{equation}\label{ADGam}
{\cal A}_{\rm \Delta\Gamma}(B_q\to f)\equiv
\frac{2\,\mbox{Re}\,\xi^{(q)}_f}{1+\bigl|\xi^{(q)}_f
\bigr|^2},
\end{equation}
which is not independent from ${\cal A}^{\mbox{{\scriptsize 
dir}}}_{\mbox{{\scriptsize CP}}}(B_q\to f)$ and 
${\cal A}^{\mbox{{\scriptsize mix}}}_{\mbox{{\scriptsize CP}}}(B_q\to f)$:
\begin{equation}\label{Obs-rel}
\Bigl[{\cal A}_{\rm CP}^{\rm dir}(B_s\to f)\Bigr]^2+
\Bigl[{\cal A}_{\rm CP}^{\rm mix}(B_s\to f)\Bigr]^2+
\Bigl[{\cal A}_{\Delta\Gamma}(B_s\to f)\Bigr]^2=1.
\end{equation}
Essentially all the information needed to evaluate the CP asymmetry
(\ref{ee6}) is included in the following quantity:
\begin{equation}
\xi_f^{(q)}=\mp\,e^{-i\phi_q}\,
\frac{A(\overline{B^0_q}\to f)}{A(B^0_q\to f)}=
\mp\,e^{-i\phi_q}\,
\frac{\sum\limits_{j=u,c}V_{jr}^\ast V_{jb}\,
{\cal M}^{jr}}{\sum\limits_{j=u,c}V_{jr}V_{jb}^\ast\,
{\cal M}^{jr}}\,,
\end{equation}
where the ${\cal M}^{jr}$ denote hadronic matrix elements 
of certain four-quark operators, $r\in\{d,s\}$ distinguishes 
between $\bar b\to\bar d$ and $\bar b\to\bar s$ transitions, and 
\begin{equation}
\phi_q=\left\{\begin{array}{cr}
+2\beta&\mbox{($q=d$)}\\
-2\delta\gamma&\mbox{($q=s$)}\end{array}\right.
\end{equation}
is the weak $B_q^0$--$\overline{B_q^0}$ mixing phase. In general, the 
observable $\xi_f^{(q)}$ suffers from hadronic uncertainties, which are 
due to the hadronic matrix elements ${\cal M}^{jr}$. However, if the 
decay $B_q\to f$ is dominated by a single CKM amplitude, the corresponding 
matrix elements cancel, and $\xi_f^{(q)}$ takes the simple form
\begin{equation}\label{ee10}
\xi_f^{(q)}=\mp\exp\left[-i\left(\phi_q-\phi_{\mbox{{\scriptsize 
D}}}^{(f)}\right)\right],
\end{equation}
where $\phi_{\mbox{{\scriptsize D}}}^{(f)}$ is a weak decay phase, 
which is given by
\begin{equation}
\phi_{\mbox{{\scriptsize D}}}^{(f)}=\left\{\begin{array}{cc}
-2\gamma&\mbox{for dominant 
$\bar b\to\bar u\,u\,\bar r$ CKM amplitudes,}\\
0&\,\mbox{for dominant $\bar b\to\bar c\,c\,\bar r\,$ CKM 
amplitudes.}
\end{array}\right.
\end{equation}

\boldmath
\subsection{The ``Gold-Plated'' Mode 
$B_d\to J/\psi\, K_{\rm S}$}\label{sec:BdPsiKS}
\unboldmath

Probably the most important application of the formalism discussed in the 
previous subsection is the decay $B_d\to J/\psi\, 
K_{\mbox{{\scriptsize S}}}$, which is dominated by the 
$\bar b\to\bar c\,c\,\bar s$ CKM amplitude \cite{rev}, implying
\begin{equation}\label{e12}
{\cal A}^{\mbox{{\scriptsize mix--ind}}}_{\mbox{{\scriptsize
CP}}}(B_d\to J/\psi\, K_{\mbox{{\scriptsize S}}})=+\sin[-(2\beta-0)]\,.
\end{equation}
Since (\ref{ee10}) applies with excellent accuracy to $B_d\to J/\psi\, 
K_{\mbox{{\scriptsize S}}}$ -- the point is that penguins
enter essentially with the same weak phase as the leading tree
contribution, as is discussed in more detail in 
Subsection~\ref{sec:BsPsiKS} -- it is referred to as the 
``gold-plated'' mode to determine the CKM angle $\beta$ \cite{bisa}.
Strictly speaking, mixing-induced CP violation in $B_d\to J/\psi\, K_{\rm S}$ 
probes $\sin(2\beta+\phi_K)$, where $\phi_K$ is related to the CP-violating 
weak $K^0$--$\overline{K^0}$ mixing phase. Similar modifications of 
(\ref{ee10}) and of the corresponding CP asymmetries must also be 
performed for other final-state configurations containing
$K_{\rm S}$- or $K_{\rm L}$-mesons. However, $\phi_K$ is negligibly small in 
the Standard Model, and -- owing to the small value of the CP-violating 
parameter $\varepsilon_K$ of the neutral kaon system -- can only be 
affected by very contrived models of new physics \cite{nir-sil}.

First attempts to measure $\sin(2\beta)$ through the CP asymmetry 
(\ref{e12}) have recently been performed by the OPAL and CDF collaborations 
\cite{sin2b-exp}:
\begin{equation}
\sin(2\beta)=\left\{\begin{array}{ll}
3.2^{+1.8}_{-2.0}\pm0.5&\mbox{(OPAL Collaboration),}\\
0.79^{+0.41}_{-0.44}&\,\mbox{\,(CDF\, Collaboration).}
\end{array}\right.
\end{equation}
Although the experimental uncertainties are very large, it is interesting 
to note that these results favour the Standard-Model expectation of a 
{\it positive} value of $\sin(2\beta)$. In the $B$-factory era, an 
experimental uncertainty of $\left.\Delta\sin(2\beta)\right|_{\rm exp}=0.08$ 
seems to be achievable, whereas second-generation experiments of the LHC 
era aim at $\left.\Delta\sin(2\beta)\right|_{\rm exp}={\cal O}(0.01)$.

Another important implication of the Standard Model, which is 
interesting for the search of new physics, is the following relation: 
\begin{equation}\label{e13}
{\cal A}^{\mbox{{\scriptsize dir}}}_{\mbox{{\scriptsize
CP}}}(B_d\to J/\psi\, K_{\mbox{{\scriptsize S}}})\approx0\approx
{\cal A}_{\mbox{{\scriptsize CP}}}(B^\pm\to J/\psi\, K^\pm).
\end{equation}
In view of the tremendous accuracy that can be achieved in the LHC era, 
it is an important issue to investigate the theoretical accuracy of 
(\ref{e12}) and (\ref{e13}). A very interesting
channel in this respect is $B_s\to J/\psi\,K_{\rm S}$ \cite{BsPsiK}, 
allowing us to extract $\gamma$ and to control the -- presumably very 
small -- penguin uncertainties in the determination of $\beta$ from the 
CP-violating effects in $B_d\to J/\psi\,K_{\rm S}$. We shall come back to
this strategy in Subsection~\ref{sec:BsPsiKS}. 

\boldmath
\subsection{The Decay $B_d\to\pi^+\pi^-$}
\unboldmath

If this mode would not receive penguin contributions, its mixing-induced
CP asymmetry would allow a measurement of $\sin(2\alpha)$:
\begin{equation}
{\cal A}^{\mbox{{\scriptsize mix--ind}}}_{\mbox{{\scriptsize
CP}}}(B_d\to\pi^+\pi^-)=-\sin[-(2\beta+2\gamma)]=-\sin(2\alpha).
\end{equation}
However, this relation is strongly affected by penguin effects, 
which were analysed by many authors \cite{alpha-uncert,charles}. There 
are various methods on the market to control the corresponding hadronic
uncertainties. Unfortunately, these strategies are usually rather 
challenging from an experimental point of view. 

The best known approach was proposed by Gronau and London \cite{gl}. 
It makes use of the $SU(2)$ isospin relation
\begin{equation}
\sqrt{2}\,A(B^+\to\pi^+\pi^0)=A(B^0_d\to\pi^+\pi^-)+
\sqrt{2}\,A(B^0_d\to\pi^0\pi^0)
\end{equation}
and of its CP-conjugate, which can be represented in the complex plane as 
two triangles. The sides of these triangles can be determined through the 
corresponding branching ratios, while their relative orientation can be 
fixed by measuring the CP-violating observable ${\cal A}^{\mbox{{\scriptsize 
mix--ind}}}_{\mbox{{\scriptsize CP}}}(B_d\to\pi^+\pi^-)$ \cite{rev}. 
Following these lines, it is in principle possible to take into account the 
QCD penguin effects in the extraction of $\alpha$. It should be noted that 
EW penguins cannot be controlled with the help of this isospin strategy. 
However, their effect is expected to be rather small, and -- as was pointed 
out recently \cite{BF,GPY} -- can be included through an additional 
theoretical input. Unfortunately, the Gronau--London approach suffers from an 
experimental problem, since the measurement of $\mbox{BR}(B_d\to\pi^0\pi^0)$,
which is expected to be at most of ${\cal O}(10^{-6})$, is very difficult. 
However, upper bounds on the CP-averaged $B_d\to\pi^0\pi^0$ branching 
ratio may already be useful to put upper bounds on the QCD penguin uncertainty 
that affects the determination of $\alpha$ \cite{charles,gq-alpha}.

Alternative methods to control the penguin uncertainties in the 
extraction of $\alpha$ from $B_d\to\pi^+\pi^-$ are very desirable. An 
important one for the asymmetric $e^+$--$e^-$ $B$-factories is provided 
by $B\to\rho\,\pi$ modes \cite{Brhopi}. Here the isospin triangle relations 
are replaced by pentagonal relations, and the corresponding approach is 
rather complicated. As we will see in Subsection~\ref{sec:BsKK}, an 
interesting strategy for second-generation $B$-physics experiments at 
hadron machines to make use of the CP-violating observables of 
$B_d\to\pi^+\pi^-$ is offered by the mode $B_s\to K^+K^-$, allowing a 
simultaneous determination of $\beta$ and $\gamma$ {\it without} any 
assumptions about penguin topologies \cite{BsKK}. 

The observation of $B_d\to\pi^+\pi^-$ has very recently been announced 
by the CLEO collaboration, with a branching ratio of
$0.47^{+0.18}_{-0.15}\pm0.13$ \cite{cleo-Bpipi}. Other CLEO results
on $B\to\pi K$ modes (see Subsection~\ref{sec:BpiK}) indicate that QCD 
penguins play in fact an important role, and that we definitely have to 
worry about them in the extraction of $\alpha$ from $B_d\to\pi^+\pi^-$. 
Needless to note that also a better theoretical understanding of the 
hadronization dynamics of $B_d\to\pi^+\pi^-$ would be very helpful in 
this respect. In a recent paper \cite{BBNS}, an interesting step towards 
this goal was performed.

\boldmath
\subsection{Extracting $2\beta+\gamma$ from $B_d\to D^{(\ast)\pm}\pi^\mp$ 
Decays}\label{sec:BDpi}
\unboldmath

\begin{figure}
\vskip -0.5truein
\begin{center}
\leavevmode
\epsfysize=4truecm 
\epsffile{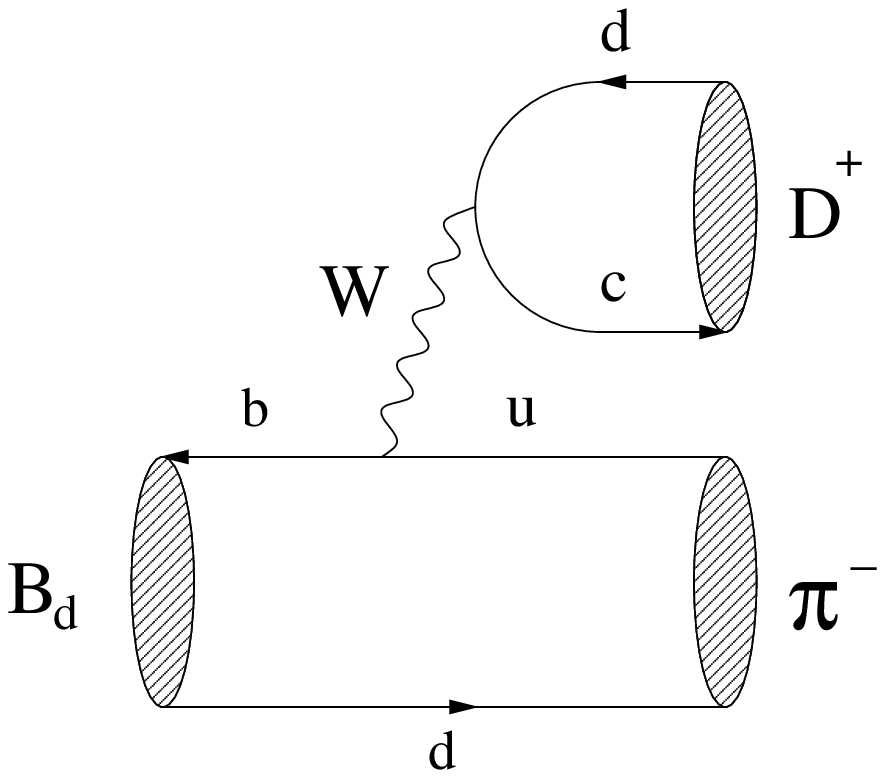} \hspace*{1truecm}
\epsfysize=4truecm 
\epsffile{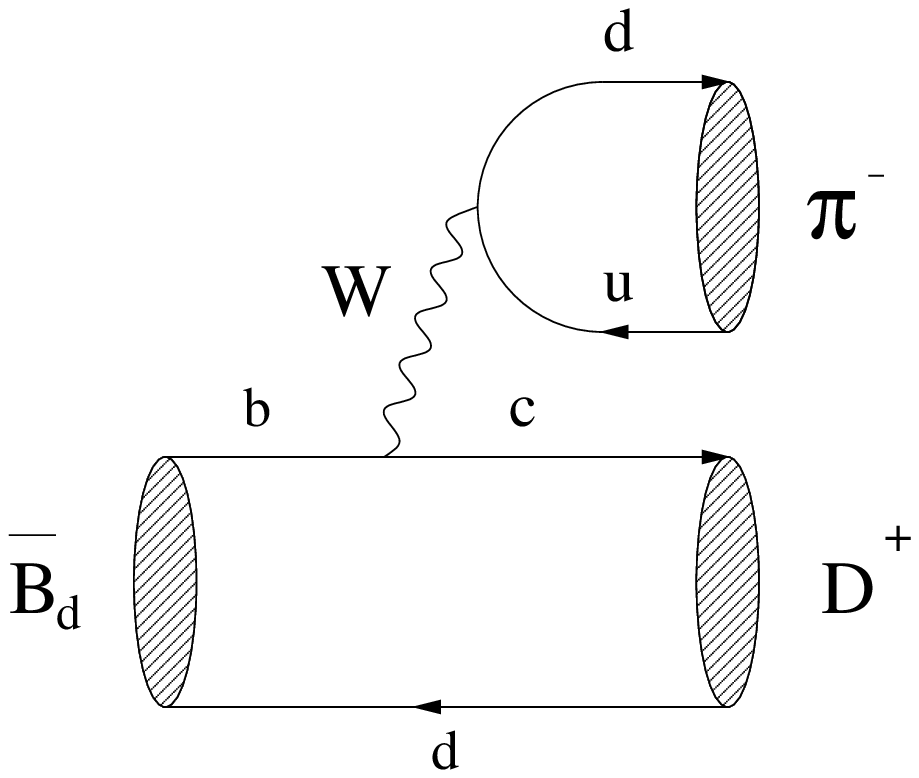}
\end{center}
\caption[]{Feynman diagrams contributing to $B^0_d, \overline{B^0_d}\to 
D^{(\ast)+}\pi^-$.}\label{fig:BDpi}
\end{figure}

The final states of the pure ``tree'' decays $B_d\to D^{(\ast)\pm}\pi^\mp$
are not CP eigenstates. However, as can be seen in Fig.~\ref{fig:BDpi}, 
$B^0_d$- and $\overline{B^0_d}$-mesons may both decay into the
$D^{(\ast)+}\pi^-$ final state, thereby leading to interference effects 
between $B^0_d$--$\overline{B^0_d}$ mixing and decay processes. Consequently, 
the time-dependent decay rates for initially, i.e.\ at time $t=0$, present 
$B^0_d$- or $\overline{B^0_d}$-mesons decaying into the final state 
$f\equiv D^{(\ast)+}\pi^-$ allow us to determine the observable \cite{rev}
\begin{equation}
\xi_f^{(d)}=-\,e^{-i\phi_d}
\frac{A(\overline{B^0_d}\to f)}{A(B^0_d\to f)}
=-\,e^{-i(\phi_d+\gamma)}\frac{1}{\lambda^2R_b}
\frac{\overline{M}_f}{M_{\overline f}},
\end{equation}
whereas those corresponding to $\bar f\equiv D^{(\ast)-}\pi^+$ allow us 
to extract
\begin{equation}
\xi_{\bar f}^{(d)}=-\,e^{-i\phi_d}
\frac{A(\overline{B^0_d}\to \bar f)}{A(B^0_d\to \bar f)}=
-\,e^{-i(\phi_d+\gamma)}\lambda^2R_b\,\frac{M_{\overline f}}{\overline{M}_f}.
\end{equation}
Here $R_b\equiv|V_{ub}/(\lambda V_{cb})|=0.41\pm0.07$ is the usual 
CKM factor, and
\begin{eqnarray}
\overline{M}_f&\equiv&\Bigl\langle f\Bigl|\overline{O}_1(\mu){\cal C}_1(\mu)+
\overline{O}_2(\mu){\cal C}_2(\mu)\Bigr|\overline{B^0_d}\Bigr\rangle\\
M_{\overline{f}}&\equiv&\Bigl\langle\overline{f}\Bigl|O_1(\mu){\cal C}_1(\mu)+
O_2(\mu){\cal C}_2(\mu)\Bigr|\overline{B^0_d}\Bigr\rangle
\end{eqnarray}
are hadronic matrix elements of the following current--current operators:
\begin{equation}
\begin{array}{rclrcl}
\overline{O}_1&=&(\bar d_\alpha u_\beta)_{\mbox{{\scriptsize 
V--A}}}\left(\bar c_\beta b_\alpha\right)_{\mbox{{\scriptsize V--A}}},&
\overline{O}_2&=&(\bar d_\alpha u_\alpha)_{\mbox{{\scriptsize 
V--A}}}\left(\bar c_\beta b_\beta\right)_{\mbox{{\scriptsize V--A}}},\\
O_1&=&(\bar d_\alpha c_\beta)_{\mbox{{\scriptsize V--A}}}
\left(\bar u_\beta b_\alpha\right)_{\mbox{{\scriptsize V--A}}},&
O_2&=&(\bar d_\alpha c_\alpha)_{\mbox{{\scriptsize V--A}}}
\left(\bar u_\beta b_\beta\right)_{\mbox{{\scriptsize 
V--A}}}.
\end{array}
\end{equation}
The observables $\xi_f^{(d)}$ and $\xi_{\bar f}^{(d)}$ allow a
{\it theoretically clean} extraction of the weak phase $\phi_d+\gamma$ 
\cite{BDpi}, as the hadronic matrix elements $\overline{M}_f$ and 
$M_{\overline{f}}$ cancel in 
\begin{equation}\label{Prod}
\xi_f^{(d)}\times\xi_{\bar f}^{(d)}=e^{-2i(\phi_d+\gamma)}.
\end{equation}
Since the $B^0_d$--$\overline{B^0_d}$ mixing phase $\phi_d$, i.e.\ $2\beta$,
can be determined rather straightforwardly with the help of the 
``gold-plated'' mode $B_d\to J/\psi\, K_{\rm S}$, we may extract the 
CKM angle $\gamma$ from (\ref{Prod}). As the $\bar b\to\bar u$ 
quark-level transition in Fig.~\ref{fig:BDpi} is doubly Cabibbo-suppressed by 
$\lambda^2R_b\approx0.02$ with respect to the $b\to c$ transition, the 
interference effects are tiny. However, the branching ratios 
are large (${\cal O}(10^{-3})$), and the $D^{(\ast)\pm}\pi^\mp$ states can 
be reconstructed with a good efficiency and modest backgrounds. Consequently, 
$B_d\to D^{(\ast)\pm}\pi^\mp$ decays offer an interesting strategy to
determine $\gamma$ \cite{BDpi-exp}. For the most optimistic scenario,
an accuracy of $\gamma$ at the level of $4^\circ$ may be achievable at 
LHCb after 5 years of taking data. 

\boldmath
\subsection{The ``El Dorado'' for Hadron Machines: $B_s$ System}
\unboldmath

Since the $e^+$--\,$e^-$ $B$-factories operating at the $\Upsilon(4S)$ 
resonance will not be in a position to explore the $B_s$ system, it
is of particular interest for hadron machines. There are important 
differences to the $B_d$ system:
\begin{itemize}
\item Within the Standard Model, a large $B^0_s$--$\overline{B^0_s}$ 
mixing parameter $x_s\equiv \Delta M_s/\Gamma_s={\cal O}(20)$ is expected, 
whereas the mixing phase $\phi_s=-2\lambda^2\eta$ is expected to be 
very small.

\item There may be a sizeable width difference $\Delta\Gamma_s\equiv
\Gamma_{\rm H}^{(s)}-\Gamma_{\rm L}^{(s)}$; the most recent theoretical 
analysis gives $\Delta\Gamma_s/\Gamma_s={\cal O}(10\%)$ \cite{DGamma-cal}.
\end{itemize}

\noindent There is an interesting correlation between $\Delta\Gamma_s$ and 
$\Delta M_s$:
\begin{equation}
\frac{\Delta\Gamma_s}{\Gamma_s}\approx-\,\frac{3\pi}{2S(x_t)}\,
\frac{m_b^2}{M_W^2}\,\frac{\Delta M_s}{\Gamma_s},
\end{equation}
where $S(x_t)$ denotes one of the well-known Inami--Lim functions. The
present experimental lower limit on $\Delta M_s$ is given by 
$\Delta M_s>12.4\,\mbox{ps}^{-1}$ (95\% C.L.). Interestingly, this lower
bound already puts constraints on the allowed region for the apex of the
unitarity triangle shown in Fig.\ \ref{fig:UT}\,(a). A detailed discussion
of this feature can be found, for instance, in \cite{BF-rev}. 

It is also interesting to note that the non-vanishing width difference 
$\Delta\Gamma_s$ may allow studies of CP-violating effects in 
``untagged'' $B_s$ rates \cite{dun,FD}:
\begin{equation}
\Gamma[f(t)]\equiv\Gamma(B^0_s(t)\to f)+\Gamma(\overline{B^0_s}(t)
\to f)\propto R_{\rm L}e^{-\Gamma_{\rm L}^{(s)}t}+
R_{\rm H}e^{-\Gamma_{\rm H}^{(s)}t},
\end{equation}
where there are no rapid oscillatory $\Delta M_st$ terms present. 
Studies of such untagged rates, allowing us to extract the observable 
${\cal A}_{\Delta\Gamma}$ introduced in (\ref{ADGam}) through
\begin{equation}
{\cal A}_{\Delta\Gamma}=\frac{R_{\rm H}-R_{\rm L}}{R_{\rm H}+R_{\rm L}},
\end{equation}
are more promising than ``tagged'' rates in terms of efficiency, acceptance 
and purity. Let us next have a brief look at the $B_s$ benchmark modes 
to extract CKM phases.

\boldmath
\subsubsection{$B_s\to D_s^\pm K^\mp$}
\unboldmath

These decays, which receive only contributions from tree-diagram-like
topologies, are the $B_s$ counterparts of the $B_d\to D^{(\ast)\pm}\pi^\mp$
modes discussed in Subsection~\ref{sec:BDpi}, and probe the CKM combination 
$\gamma-2\delta\gamma$ instead of $\gamma+2\beta$ in a {\it theoretically 
clean} way \cite{adk}. Since one decay path is only suppressed by 
$R_b\approx0.41$, and is not doubly Cabibbo-suppressed by $\lambda^2 R_b$,
as in $B_d\to D^{(\ast)\pm}\pi^\mp$, the interference effects in 
$B_s\to D_s^\pm K^\mp$ are much larger.

\boldmath
\subsubsection{$B_s\to J/\psi\,\phi$}
\unboldmath

The decay $B_s\to J/\psi[\to l^+l^-]\,\phi[\to K^+K^-]$ is the $B_s$ 
counterpart of the ``gold-plated'' mode $B_d\to J/\psi\,K_{\rm S}$. The 
observables of the angular distribution of its decay products provide
interesting strategies to extract the $B_s^0$--$\overline{B_s^0}$ mixing 
parameters $\Delta M_s$ and $\Delta\Gamma_s$, as well as the CP-violating
weak mixing phase $\phi_s\equiv-2\delta\gamma$ \cite{ddf1}. Because of
$\delta\gamma=\lambda^2\eta$, this phase would allow us to extract the
Wolfenstein parameter $\eta$. However, since $\delta\gamma={\cal O}(0.02)$ 
is tiny within the Standard Model, its extraction from the 
$B_s\to J/\psi\,\phi$ angular distribution may well be sizeably affected 
by penguin topologies. These uncertainties, which are an important issue 
for second-generation $B$-physics experiments at hadron machines, can be 
controlled with the help of the decay $B_d\to J/\psi\,\rho^0$ \cite{RF-ang},
as is discussed in more detail in Subsection~\ref{sec:ang}. 

Since the CP-violating effects in $B_s\to J/\psi\,\phi$ are very small in 
the Standard Model, they provide an interesting probe for new physics 
\cite{nir-sil}. In the case of $B_s\to J/\psi\,\phi$, the preferred mechanism 
for new physics to manifest itself in the corresponding observables are 
CP-violating new-physics contributions to  $B^0_s$--$\overline{B^0_s}$ mixing. 
In various scenarios for new physics, for example in the left--right-symmetric 
model with spontaneous CP violation \cite{bbmr}, there are in fact large 
contributions to the $B_s^0$--$\overline{B_s^0}$ mixing phase. 

Because of its very favourable experimental signature, studies of 
$B_s\to J/\psi\,\phi$ are not only promising for dedicated second-generation 
$B$-physics experiments, such as LHCb or BTeV, but also for ATLAS and CMS 
\cite{smizanska}.

\boldmath
\subsection{CP Violation in Charged $B$ Decays}
\unboldmath

Since there are no mixing effects present in the charged $B$-meson system, 
non-vanishing CP asymmetries of the kind 
\begin{equation}\label{CP-charged}
{\cal A}_{\mbox{{\scriptsize CP}}}\equiv\frac{\Gamma(B^+\to\overline{f})-
\Gamma(B^-\to f)}{\Gamma(B^+\to\overline{f})+\Gamma(B^-\to f)}
\end{equation}
would give us unambiguous evidence for ``direct'' CP violation in the 
$B$ system, which has recently been demonstrated in the kaon system by 
the new experimental results of the KTeV (Fermilab) and NA48 (CERN) 
collaborations for Re$(\varepsilon'/\varepsilon)$ \cite{calvetti}. 

The CP asymmetries (\ref{CP-charged}) arise from the interference between 
decay amplitudes with both different CP-violating weak and different 
CP-conserving strong phases. In the Standard Model, the weak phases are 
related to the phases of the CKM matrix elements, whereas the strong phases 
are induced by final-state-interaction processes. In general, the strong 
phases introduce severe theoretical uncertainties into the calculation of 
${\cal A}_{\mbox{{\scriptsize CP}}}$, thereby destroying the clean 
relation to the CP-violating weak phases. However, there is an important 
tool to overcome these problems, which is provided by {\it amplitude 
relations} between certain non-leptonic $B$ decays. There are two kinds 
of such relations:
\begin{itemize}
\item Exact relations: $B\to DK$  (pioneered by Gronau and Wyler
\cite{gw}).
\item Approximate relations, based on flavour-symmetry arguments and certain 
plausible dynamical assumptions: $B\to \pi K$, $\pi\pi$, $K\overline{K}$ 
(pioneered by Gronau, Hern\'andez, London and Rosner \cite{GRL,GHLR}).
\end{itemize}
Unfortunately, the $B\to DK$ approach, which allows a {\it theoretically 
clean} determination of $\gamma$, involves amplitude triangles that are 
expected to be very squashed. Moreover, we have to deal with additional 
experimental problems~\cite{ads}, so that 
this approach is very challenging from a practical point of view. More 
refined variants were proposed in \cite{ads}. Let us note that the 
colour-allowed decay $B^-\to D^0K^-$ was observed by CLEO 
in 1998 \cite{cleo-bdk}.

The flavour-symmetry relations between the $B\to \pi K$, $\pi\pi$, 
$K\overline{K}$ decay amplitudes have received considerable attention in 
the literature during the last couple of years and led to interesting 
strategies to probe the CKM angle $\gamma$, which are the subject of the 
following subsection.

\section{A Closer Look at New Strategies to\\
Extract CKM Phases}\label{sec:New-Strat}

\boldmath
\subsection{Extracting $\gamma$ from $B\to\pi K$ Decays}\label{sec:BpiK}
\unboldmath

In order to obtain direct information on $\gamma$ in an experimentally 
feasible way, $B\to\pi K$ decays seem very promising. Fortunately, 
experimental data on these modes are now starting to become available. 
In 1997, the CLEO collaboration reported the first results on the decays 
$B^\pm\to\pi^\pm K$ and $B_d\to\pi^\mp K^\pm$; in the following year, the 
first observation of $B^\pm\to\pi^0K^\pm$ was announced. So far, only results 
for CP-averaged branching ratios have been reported, with values at the 
$10^{-5}$ level and large experimental uncertainties \cite{berkelman}. 
However, already such CP-averaged branching ratios may lead to highly 
non-trivial constraints on $\gamma$ \cite{FM}. So far, the following three 
combinations of $B\to\pi K$ decays were considered in the literature: 
$B^\pm\to\pi^\pm K$ and $B_d\to\pi^\mp K^\pm$ \cite{FM}--\cite{GroRo}, 
$B^\pm\to\pi^\pm K$ and $B^\pm\to\pi^0 K^\pm$ \cite{BF,GRL,NR}, as well 
as the combination of the neutral decays $B_d\to\pi^0 K$ and 
$B_d\to\pi^\mp K^\pm$ \cite{BF}. 

\boldmath
\subsubsection{The $B^\pm\to\pi^\pm K$, $B_d\to\pi^\mp K^\pm$ Strategy}
\unboldmath

Within the framework of the Standard Model, the most important contributions
to these decays originate from QCD penguin topologies. Making use of the 
$SU(2)$ isospin symmetry of strong interactions, we obtain
\begin{equation}\label{rel1}
A(B^+\to\pi^+K^0)\equiv P,\quad A(B_d^0\to\pi^-K^+)=-\,
\left[P+T+P_{\rm ew}^{\rm C}\right],
\end{equation}
where 
\begin{equation}
T\equiv|T|e^{i\delta_T}e^{i\gamma} \quad\mbox{and}\quad
P_{\rm ew}^{\rm C}\equiv-\,\left|P_{\rm ew}^{\rm C}\right|
e^{i\delta_{\rm ew}^{\rm C}}
\end{equation}
are due to tree-diagram-like topologies and EW penguins, respectively. 
The label ``C'' reminds us that only ``colour-suppressed''
EW penguin topologies contribute to $P_{\rm ew}^{\rm C}$. Making use of 
the unitarity of the CKM matrix and applying the Wolfenstein parametrization,
generalized to include non-leading terms in $\lambda$ \cite{BLO}, 
we obtain \cite{defan}
\begin{equation}
P\equiv A(B^+\to\pi^+K^0)=-\left(1-\frac{\lambda^2}{2}\right)\lambda^2A\left[
1+\rho\,e^{i\theta}e^{i\gamma}\right]{\cal P}_{tc}\,,
\end{equation}
where
\begin{equation}
\rho\,e^{i\theta}=\frac{\lambda^2R_b}{1-\lambda^2/2}
\left[1-\left(\frac{{\cal P}_{uc}+{\cal A}}{{\cal P}_{tc}}\right)\right].
\end{equation}
Here ${\cal P}_{tc}\equiv|{\cal P}_{tc}|e^{i\delta_{tc}}$ and ${\cal P}_{uc}$
describe differences of penguin topologies with internal top- and charm-quark
and up- and charm-quark exchanges, respectively, and ${\cal A}$ is due to 
annihilation topologies. It is important to note that $\rho$ is strongly 
CKM-suppressed by $\lambda^2R_b\approx0.02$. In the parametrization of the 
$B^\pm\to \pi^\pm K$ and $B_d\to\pi^\mp K^\pm$ observables, it turns out 
to be useful to introduce 
\begin{equation}
r\equiv\frac{|T|}{\sqrt{\langle|P|^2\rangle}}\,,\quad\epsilon_{\rm C}\equiv
\frac{|P_{\rm ew}^{\rm C}|}{\sqrt{\langle|P|^2\rangle}}\,,
\end{equation}
with $\langle|P|^2\rangle\equiv(|P|^2+|\overline{P}|^2)/2$, as well
as the strong phase differences
\begin{equation}
\delta\equiv\delta_T-\delta_{tc}\,,\quad\Delta_{\rm C}\equiv
\delta_{\rm ew}^{\rm C}-\delta_{tc}\,.
\end{equation}
In addition to the ratio 
\begin{equation}\label{Def-R}
R\equiv\frac{\mbox{BR}(B^0_d\to\pi^-K^+)+
\mbox{BR}(\overline{B^0_d}\to\pi^+K^-)}{\mbox{BR}(B^+\to\pi^+K^0)
+\mbox{BR}(B^-\to\pi^-\overline{K^0})}
\end{equation}
of CP-averaged branching ratios, also the ``pseudo-asymmetry'' 
\begin{equation}
A_0\equiv\frac{\mbox{BR}(B^0_d\to\pi^-K^+)-\mbox{BR}(\overline{B^0_d}\to
\pi^+K^-)}{\mbox{BR}(B^+\to\pi^+K^0)+\mbox{BR}(B^-\to\pi^-\overline{K^0})}
\end{equation}
plays an important role in the probing of $\gamma$. Explicit expressions for 
$R$ and $A_0$ in terms of the parameters specified above are given in 
\cite{defan}.

\begin{figure}
\centerline{
\rotate[r]{
\epsfxsize=6.3truecm
{\epsffile{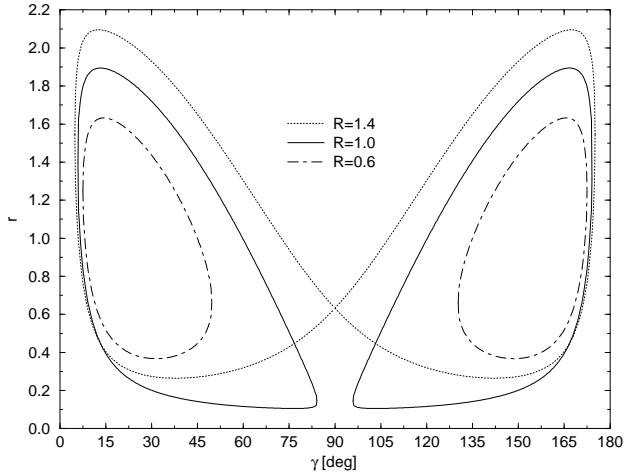}}}}
\caption[]{The contours in the $\gamma$--$r$ plane for $|A_0|=0.2$ 
($\rho=\epsilon_{\rm C}=0$).}\label{fig:g-r-cont}
\end{figure}

So far, the only available result from the CLEO collaboration is for $R$:
\begin{equation}\label{RFM-exp}
R=1.0\pm0.4,
\end{equation}
and no CP-violating effects have been reported. However, if in addition 
to $R$ also the pseudo-asymmetry $A_0$ can be measured, it is possible to 
eliminate the strong phase $\delta$ in the expression for $R$, and to 
fix contours in the $\gamma\,$--$\,r$ plane \cite{defan}. These contours, 
which are illustrated in Fig.~\ref{fig:g-r-cont}, correspond to the 
mathematical implementation of a simple triangle construction \cite{PAPIII}.
In order to determine $\gamma$, the quantity $r$, i.e.\ the magnitude 
of the ``tree'' amplitude $T$, has to be fixed. At this stage, a certain model 
dependence enters. Since the properly defined amplitude $T$ does not receive 
contributions only from colour-allowed ``tree'' topologies, but also from 
penguin and annihilation processes \cite{defan,bfm}, it may be sizeably
shifted from its ``factorized'' value. Consequently, estimates of the 
uncertainty of $r$ using the factorization hypothesis, yielding typically 
$\Delta r={\cal O}(10\%)$, may be too optimistic.

Interestingly, it is possible to derive bounds on $\gamma$ that do {\it not}
depend on $r$ at all \cite{FM}. To this end, we eliminate again $\delta$ 
in $R$ through $A_0$. If we now treat $r$ as a ``free'' variable, we find 
that $R$ takes the minimal value \cite{defan} 
\begin{equation}\label{Rmin}
R_{\rm min}=\kappa\,\sin^2\gamma\,+\,
\frac{1}{\kappa}\left(\frac{A_0}{2\,\sin\gamma}\right)^2\geq
\kappa\,\sin^2\gamma,
\end{equation}
where
\begin{equation}\label{kappa-def}
\kappa=\frac{1}{w^2}\left[\,1+2\,(\epsilon_{\rm C}\,w)\cos\Delta+
(\epsilon_{\rm C}\,w)^2\,\right],
\end{equation}
with $w=\sqrt{1+2\,\rho\,\cos\theta\cos\gamma+\rho^2}$. The inequality in 
(\ref{Rmin}) arises if we keep both $r$ and $\delta$ as free parameters 
\cite{FM}. An allowed range for $\gamma$ is related to $R_{\rm min}$, since 
values of $\gamma$ implying $R_{\rm exp}<R_{\rm min}$ are excluded. 
In particular, $A_0\not=0$ would allow us to exclude a certain range of 
$\gamma$ around $0^\circ$ or $180^\circ$, whereas a measured value of $R<1$ 
would exclude a certain range around $90^\circ$, which would be of great 
phenomenological importance. The first results reported by 
CLEO in 1997 gave $R=0.65\pm0.40$, whereas the most recent update is that 
given in (\ref{RFM-exp}). If we are willing to fix the parameter $r$, 
significantly stronger constraints on $\gamma$ can be obtained from 
$R$ \cite{BF,GPY}. In particular, these constraints require only 
$R\not=1$ and are also effective for $R>1$.

\begin{figure}
\centerline{
\epsfysize=3.7truecm
{\epsffile{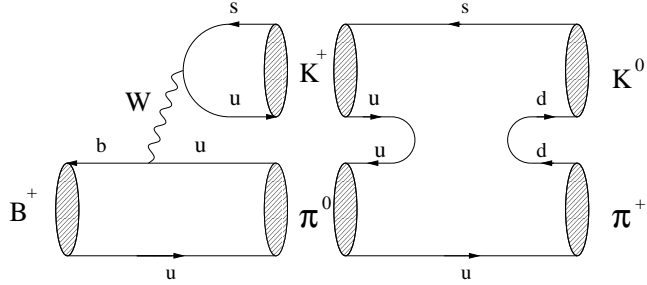}}}
\caption[]{Rescattering process contributing to 
$B^+\to\pi^+K^0$.}\label{fig:res}
\end{figure}

The theoretical accuracy of the strategies to probe $\gamma$ with the 
decays $B^\pm\to\pi^\pm K$ and $B_d\to\pi^\mp K^\pm$ is limited both 
by rescattering processes of the kind 
$B^+\to\{\pi^0K^+,\pi^0K^{\ast+},\ldots\}\to\pi^+K^0$ \cite{FSI,neubert}, 
which are illustrated in Fig.~\ref{fig:res}, and by ``colour-suppressed'' 
EW penguin contributions \cite{GroRo,neubert}. In Eq.~(\ref{Rmin}), 
these effects are described by the parameter $\kappa$. If they are 
neglected, we have $\kappa=1$. The rescattering effects, which may lead 
to values of $\rho={\cal O}(0.1)$, can be controlled in the contours in 
the $\gamma$--$r$ plane and the constraints on $\gamma$ related to
(\ref{Rmin}) through experimental data on $B^\pm\to K^\pm K$ decays, 
which are the $U$-spin counterparts of $B^\pm\to\pi^\pm K$ \cite{defan,BKK}. 
Another important indicator for large rescattering effects is provided by 
$B_d\to K^+K^-$ modes, for which there already exist stronger experimental 
constraints \cite{groro-FSI}.

An improved description of the EW penguins is possible if we use the 
general expressions for the corresponding four-quark operators, 
and perform appropriate Fierz transformations. Following these 
lines \cite{defan,neubert}, we obtain
\begin{equation}\label{EWP-expr1}
q_{\rm C}\,e^{i\omega_{\rm C}}\equiv\frac{\epsilon_{\rm C}}{r}\,
e^{i(\Delta_{\rm C}-\delta)}=0.66\times \left[\frac{0.41}{R_b}\right]
\times a_{\rm C}\,e^{i\omega_{\rm C}},
\end{equation}
where $a_{\rm C}\,e^{i\omega_{\rm C}}=a_2^{\rm eff}/a_1^{\rm eff}$ is the
ratio of certain generalized ``colour factors''. Experimental data on 
$B\to D^{(\ast)}\pi$ decays imply $a_2/a_1={\cal O}(0.25)$. However, 
``colour suppression'' in $B\to\pi K$ modes may in principle be different 
from that in $B\to D^{(\ast)}\pi$ decays, in particular in the presence of 
large rescattering effects \cite{neubert}. A first step to fix the hadronic 
parameter $a_{\rm C}\,e^{i\omega_{\rm C}}$ experimentally is provided by 
the mode $B^+\to\pi^+\pi^0$ \cite{defan}; interesting constraints were
derived in \cite{GPY,pirjol}. For a detailed discussion of the impact of 
rescattering and EW penguin effects on the strategies to probe $\gamma$ 
with $B^\pm\to\pi^\pm K$ and $B_d\to\pi^\mp K^\pm$ decays, the reader is
referred to \cite{BF,bfm,BKK}.

\boldmath
\subsubsection{The Charged $B^\pm\to \pi^\pm K$, $B^\pm\to\pi^0K^\pm$
Strategy}
\unboldmath

Several years ago, Gronau, Rosner and London proposed an interesting 
$SU(3)$ strategy to determine $\gamma$ with the help of the charged decays 
$B^{\pm}\to\pi^{\pm} K$, $\pi^0K^{\pm}$, $\pi^0\pi^{\pm}$ \cite{GRL}. 
However, as was pointed out by Deshpande and He \cite{dh}, this elegant 
approach is unfortunately spoiled by EW penguins, which play an important 
role in several non-leptonic $B$-meson decays because of the large top-quark 
mass \cite{rf-ewp}. Recently, this approach was resurrected by Neubert 
and Rosner \cite{NR}, who pointed out that the EW penguin contributions 
can be controlled in this case by using only the general expressions for 
the corresponding four-quark operators, appropriate Fierz transformations, 
and the $SU(3)$ flavour symmetry (see also \cite{PAPIII}). Since a more
detailed presentation of these strategies can be found in the contribution
by D. Pirjol to these proceedings, we will just have a brief look at 
their most interesting features.

In the case of $B^+\to\pi^+K^0$, $\pi^0K^+$, the $SU(2)$ isospin 
symmetry implies
\begin{equation}\label{charged-iso}
A(B^+\to\pi^+K^0)\,+\,\sqrt{2}\,A(B^+\to\pi^0K^+)=
-\left[(T+C)\,+\,P_{\rm ew}\right].
\end{equation}
The phase structure of this relation, which has no $I=1/2$ piece, is
completely analogous to the $B^+\to\pi^+K^0$, $B^0_d\to\pi^-K^+$ case
(see (\ref{rel1})):
\begin{equation}
T+C=|T+C|\,e^{i\delta_{T+C}}\,e^{i\gamma},\quad
P_{\rm ew}=-\,|P_{\rm ew}|e^{i\delta_{\rm ew}}\,.
\end{equation}
In order to probe $\gamma$, it is useful to introduce the following
observables \cite{BF}:
\begin{eqnarray}
R_{\rm c}&\equiv&2\left[\frac{\mbox{BR}(B^+\to\pi^0K^+)+
\mbox{BR}(B^-\to\pi^0K^-)}{\mbox{BR}(B^+\to\pi^+K^0)+
\mbox{BR}(B^-\to\pi^-\overline{K^0})}\right]\label{Rc-def}\\
A_0^{\rm c}&\equiv&2\left[\frac{\mbox{BR}(B^+\to\pi^0K^+)-
\mbox{BR}(B^-\to\pi^0K^-)}{\mbox{BR}(B^+\to\pi^+K^0)+
\mbox{BR}(B^-\to\pi^-\overline{K^0})}\right],\label{A0c-def}
\end{eqnarray}
which correspond to $R$ and $A_0$; their general expressions can be 
otained from those for $R$ and $A_0$ by making the following replacements:
\begin{equation}
r\to r_{\rm c}\equiv\frac{|T+C|}{\sqrt{\langle|P|^2\rangle}}\,, \quad
\delta\to \delta_{\rm c}\equiv\delta_{T+C}-\delta_{tc}\,,\quad
P_{\rm ew}^{\rm C}\to P_{\rm ew}.
\end{equation}
The measurement of $R_{\rm c}$ and $A_0^{\rm c}$ allows us to fix 
contours in the $\gamma$--$r_c$ plane, in complete analogy to the
$B^\pm\to\pi^\pm K$, $B_d\to\pi^\mp K^\pm$ strategy. However, the
charged $B\to\pi K$ approach has interesting advantages from a
theoretical point of view. First, the $SU(3)$ symmetry allows us to 
fix $r_c\propto|T+C|$ \cite{GRL}:
\begin{equation}\label{SU3-rel1}
T+C\approx-\,\sqrt{2}\,\frac{V_{us}}{V_{ud}}\,
\frac{f_K}{f_{\pi}}\,A(B^+\to\pi^+\pi^0)\,,
\end{equation}
where $r_c$ thus determined is -- in contrast to $r$ -- not affected by 
rescattering effects. Second, in the strict $SU(3)$ limit, we have \cite{NR}
\begin{equation}\label{SU3-rel2}
q\,e^{i\omega}\equiv\left|\frac{P_{\rm ew}}{T+C}\right|\,
e^{i(\delta_{\rm ew}-\delta_{T+C})}=0.66\times
\left[\frac{0.41}{R_b}\right],
\end{equation}
which does not -- in contrast to (\ref{EWP-expr1}) -- involve a
hadronic parameter. 

The contours in the $\gamma$--$r_c$ plane may be affected -- in analogy 
to the $B^\pm\to\pi^\pm K$, $B_d\to\pi^\mp K^\pm$ case -- by rescattering 
effects \cite{BF}. They can be taken into account with the help of 
additional data \cite{defan,BKK,FSI-recent}. The major theoretical 
advantage of the $B^+\to\pi^+K^0$, $\pi^0K^+$ strategy with respect to 
$B^\pm\to\pi^\pm K$, $B_d\to\pi^\mp K^\pm$ is that $r_c$ and 
$P_{\rm ew}/(T+C)$ can be fixed by using {\it only} $SU(3)$ arguments. 
Consequently, the theoretical accuracy is mainly limited by non-factorizable 
$SU(3)$-breaking effects. 

Let us finally note that the observable $R_{\rm c}$ -- the present
CLEO result is $R_{\rm c}=2.1\pm1.1$ -- may also imply interesting 
constraints on $\gamma$ \cite{NR}. These bounds, which are conceptually
similar to \cite{FM}, are related to the extremal values of $R_{\rm c}$ 
that arise if we keep the strong phase $\delta_{\rm c}$ as an ``unknown'', 
free parameter. As the resulting general expression is rather complicated 
\cite{BF}, let us expand it in $r_c$ \cite{NR}. If we keep only the 
leading-order terms and make use of the $SU(3)$ relation (\ref{SU3-rel2}), 
we obtain
\begin{equation}\label{Rc-expansion}
\left.R_c^{\rm ext}\right|_{\delta_c}^{\rm L.O.}=
1\,\pm\,2\,r_c\,|\cos\gamma-q|.
\end{equation}
Interestingly, there are no terms of ${\cal O}(\rho)$ present in this
expression, i.e.\ rescattering effects do not enter at this level \cite{NR}.
However, final-state-interaction processes may still have a sizeable
impact on the bounds on $\gamma$ arising from the charged $B\to\pi K$ 
decays. Several strategies to control these uncertainties were considered 
in the recent literature \cite{BF,FSI-recent}.

\boldmath
\subsubsection{The Neutral $B_d\to \pi^0 K$, $B_d\to\pi^\mp K^\pm$
Strategy}
\unboldmath

At first sight, the strategies to probe $\gamma$ that are provided by 
the observables of the neutral decays $B_d\to \pi^0 K$, $\pi^\mp K^\pm$ 
are completely analogous to the charged $B^\pm\to\pi^\pm K$, $\pi^0K^\pm$ 
case \cite{BF}, as the corresponding decay amplitudes satisfy a similar
isospin relation (see (\ref{charged-iso})). However, if we require 
that the neutral kaon be observed as a $K_{\rm S}$, we have an additional 
observable at our disposal, which is due to ``mixing-induced'' CP 
violation in $B_d\to\pi^0K_{\rm S}$ and allows us to take into account 
the rescattering effects in the extraction of $\gamma$ \cite{BF}. To this 
end, time-dependent measurements are required. The theoretical accuracy 
of the neutral strategy is only limited by non-factorizable $SU(3)$-breaking 
corrections, which affect $|T+C|$ and $P_{\rm ew}$.

\subsubsection{Some Thoughts about New Physics}

Since $B^0_q$--$\overline{B^0_q}$ mixing ($q\in\{d,s\}$) is a ``rare'' 
flavour-changing neutral-current (FCNC) process, it is very likely that 
it is significantly affected by new physics, which may act upon the mixing 
parameters $\Delta M_q$ and $\Delta\Gamma_q$ as well as on the CP-violating
mixing phase $\phi_q$. Important examples for such scenarios of
new physics are non-minimal SUSY models, left--right-symmetric models,
models with extended Higgs sectors, four generations, or $Z$-mediated 
\mbox{FCNCs}~\cite{new-phys}. Since $B_d\to J/\psi\,K_{\rm S}$ and 
$B_s\to J/\psi\,\phi$ -- the benchmark modes to measure $\phi_d$ and 
$\phi_s$ -- are governed by current--current, i.e.\ ``tree'', processes, 
new physics is expected to affect their {\it decay amplitudes} in a 
minor way. Consequently, these modes still measure $\phi_d$ and $\phi_s$.

\begin{figure}
\vspace*{-1.9truecm}
\begin{tabular}{lr}
   \epsfysize=8.8cm
   \epsffile{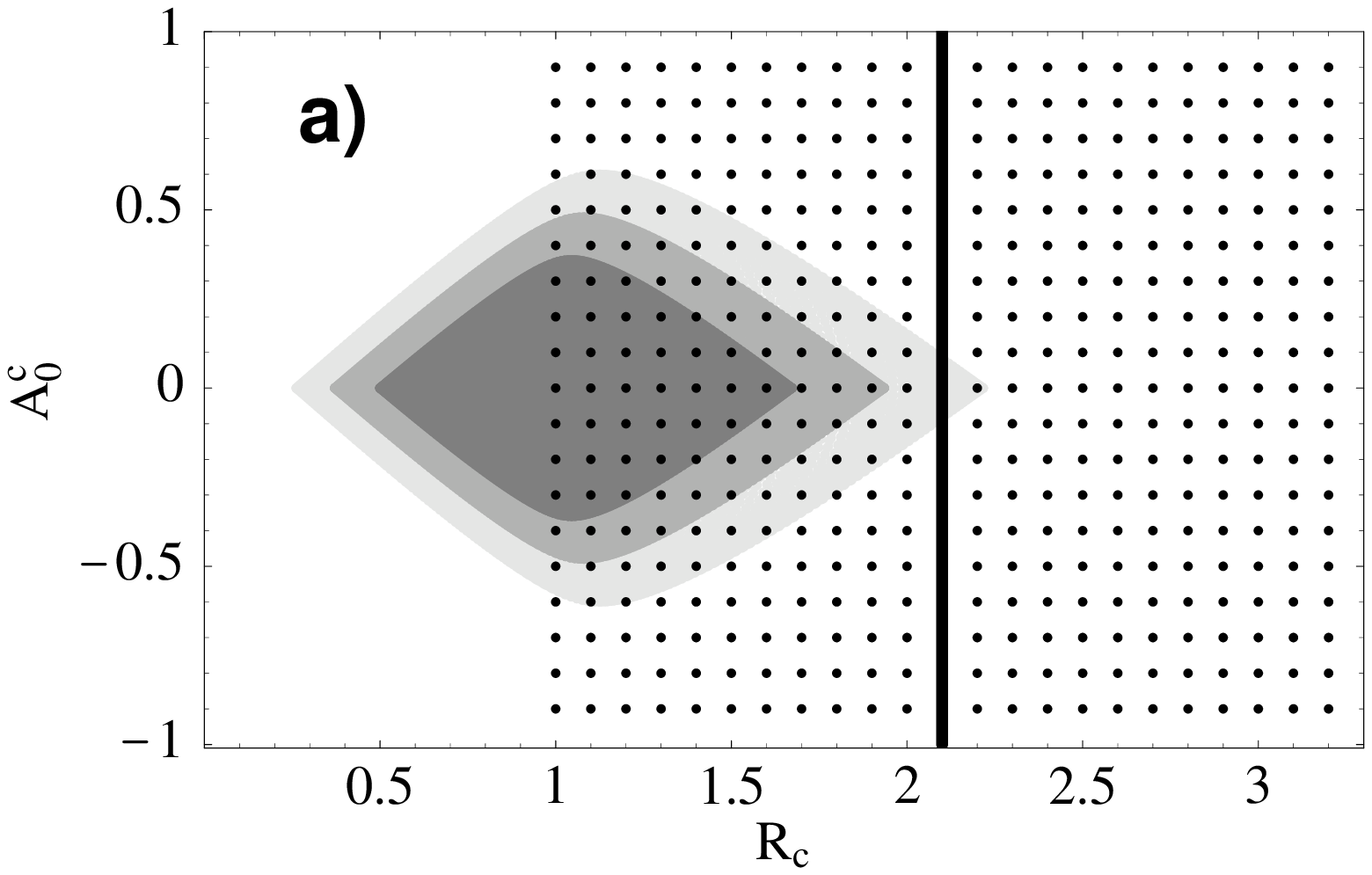}
&
   \epsfysize=8.6cm
   \epsffile{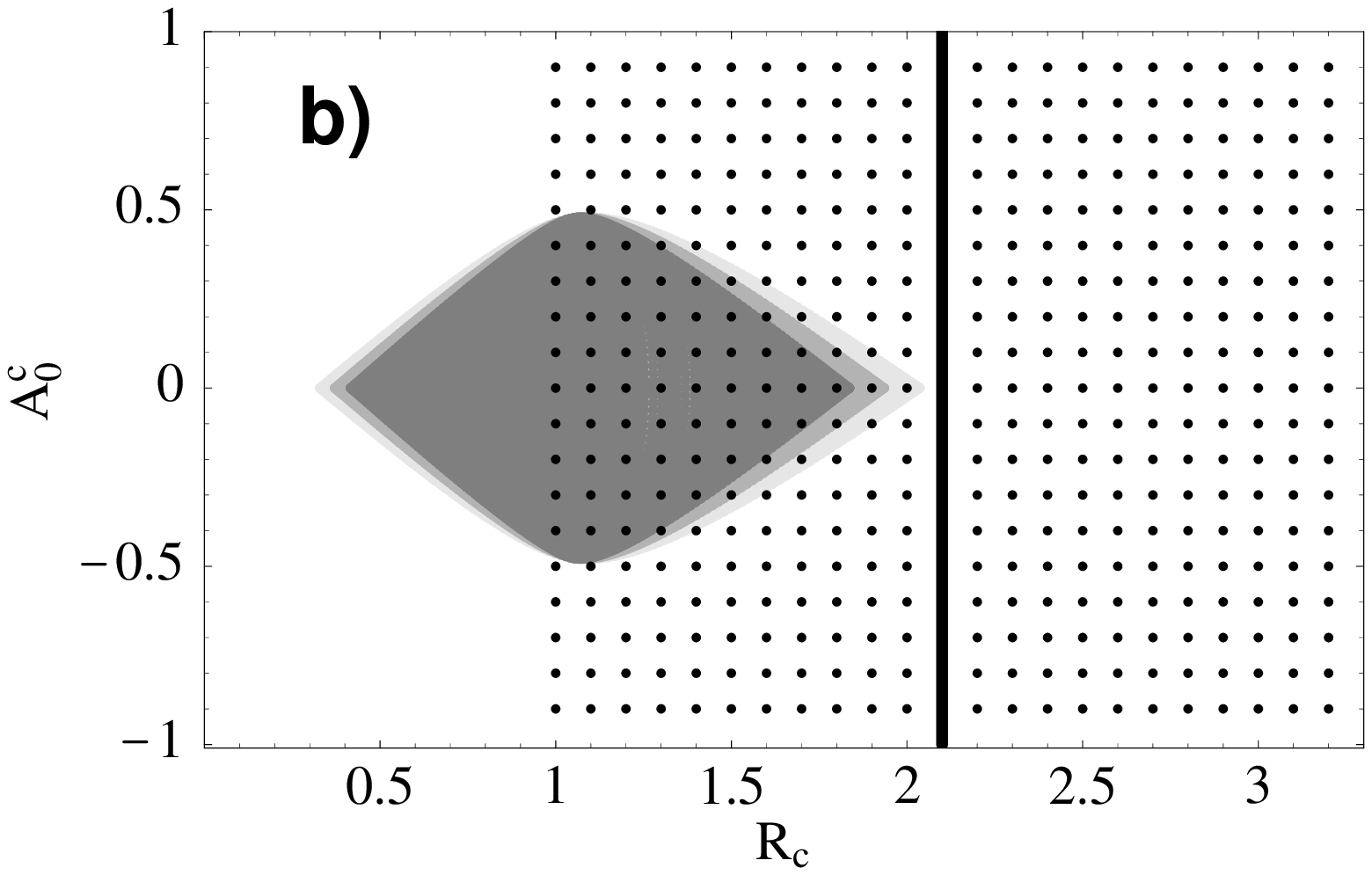}
\end{tabular}
\vspace*{-2.5truecm}
\caption[]{The allowed region in the $R_{\rm c}$--$A_0^{\rm c}$ 
plane, characterizing the $B^\pm\to\pi^\pm K$, $\pi^0K^\pm$ system
in the Standard Model: (a) $0.18\leq r_{\rm c}\leq0.30$, $q=0.63$; 
(b) $r_{\rm c}=0.24$, $0.48\leq q\leq0.78$. Rescattering effects are 
neglected.}
\label{fig:BpiK-char}
\end{figure}

\begin{figure} 
   \epsfysize=7.3cm
   \centerline{\epsffile{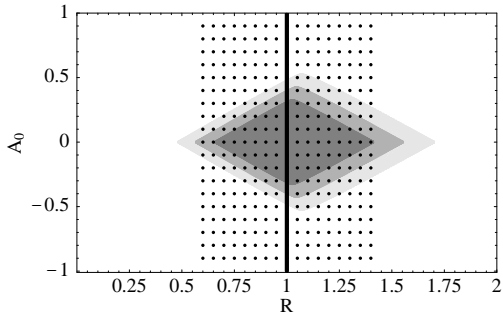}}
\vspace{-2.6cm}
\caption[]{The allowed region in the $R$--$A_0$ plane, characterizing 
the $B^\pm\to\pi^\pm K$, $B_d\to\pi^\mp K^\pm$ system within the Standard
Model for $0.16\leq r\leq0.26$, 
$q_{\rm C}\,e^{i\omega_{\rm C}}=0.66\times0.25$. Rescattering effects are 
neglected.}
\label{fig:BpiK-mix}
\end{figure}

In the clean strategies to measure $\gamma$ with the help of pure ``tree''
decays, such as $B\to DK$, $B_d\to D^{(\ast)\pm}\pi^\mp$ or 
$B_s\to D_s^\pm K^\mp$, new physics is also expected to play a very minor 
role. These strategies therefore provide a ``reference'' value for $\gamma$. 
Since, on the other hand, the $B\to\pi K$ strategies to determine $\gamma$ 
rely on the interference between tree and penguin contributions, 
discrepancies with the ``reference'' value for $\gamma$ may well show up 
in the presence of new physics. If we are lucky, we may even get immediate 
indications for new physics from $B\to\pi K$ decays \cite{FMat}, as 
the Standard Model predicts interesting correlations between the 
corresponding observables that are shown in Figs.~\ref{fig:BpiK-char} and 
\ref{fig:BpiK-mix}. Here the dotted regions correspond to the present CLEO 
results for $R_{\rm c}$ and $R$. A future measurement of observables lying 
significantly outside the allowed regions shown in these figures would 
immediately indicate the presence of new physics. Although the experimental 
uncertainties are still too large for us to draw definite conclusions, it 
is interesting to note that the present central value of $R_{\rm c}=2.1$ is 
not favoured by the Standard Model (see Fig.~\ref{fig:BpiK-char}). Moreover, 
if future measurements should stabilize at such a large value, there would 
essentially be no space left for $A^{\rm c}_0$. These features should 
be compared with the situation in Fig.~\ref{fig:BpiK-mix}. The strategies 
discussed in the following subsections are also well suited to search 
for new physics.

\boldmath
\subsection{Extracting $\gamma$ from 
$B_{s(d)}\to J/\psi\, K_{\rm S}$}\label{sec:BsPsiKS}
\unboldmath

\begin{figure}
\vskip -0.5truein
\begin{center}
\leavevmode
\epsfysize=4.5truecm 
\epsffile{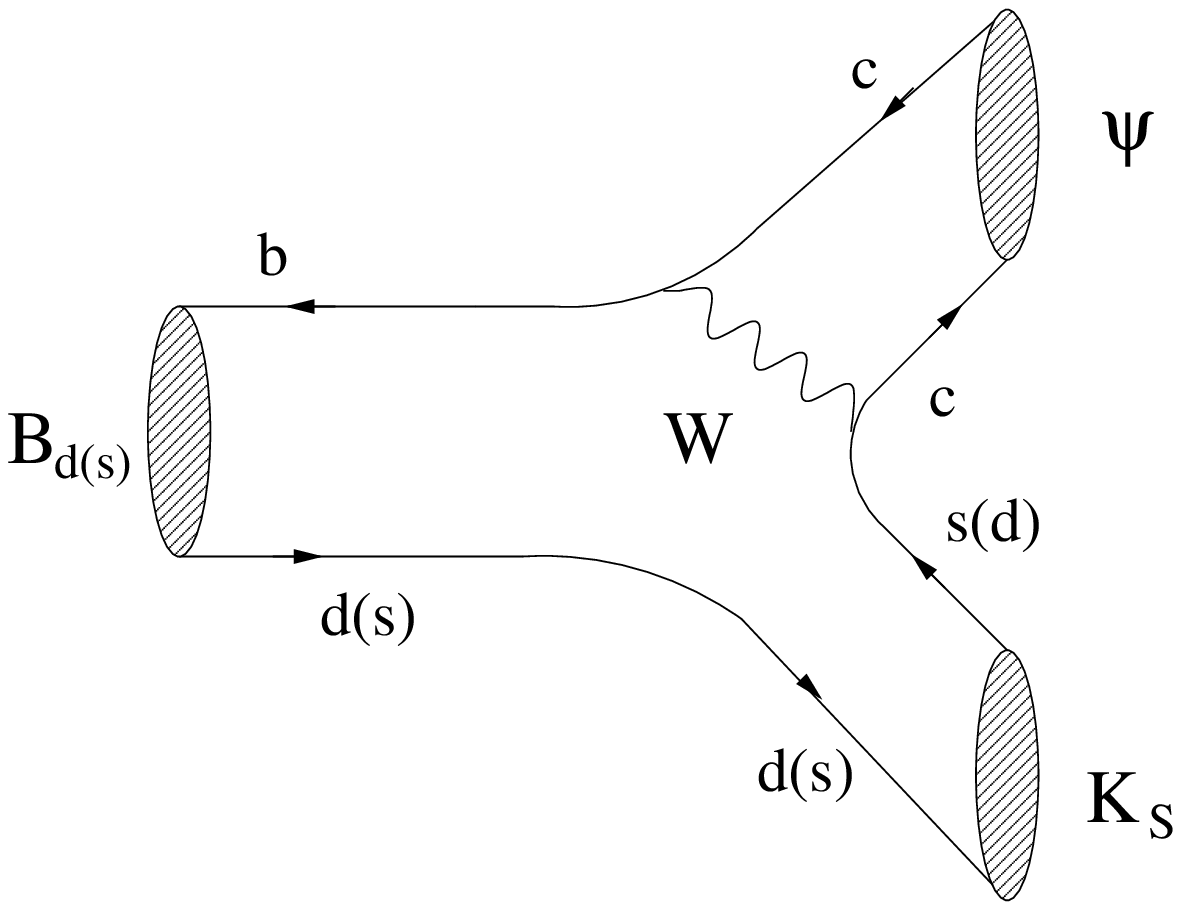} \hspace*{1truecm}
\epsfysize=4.5truecm 
\epsffile{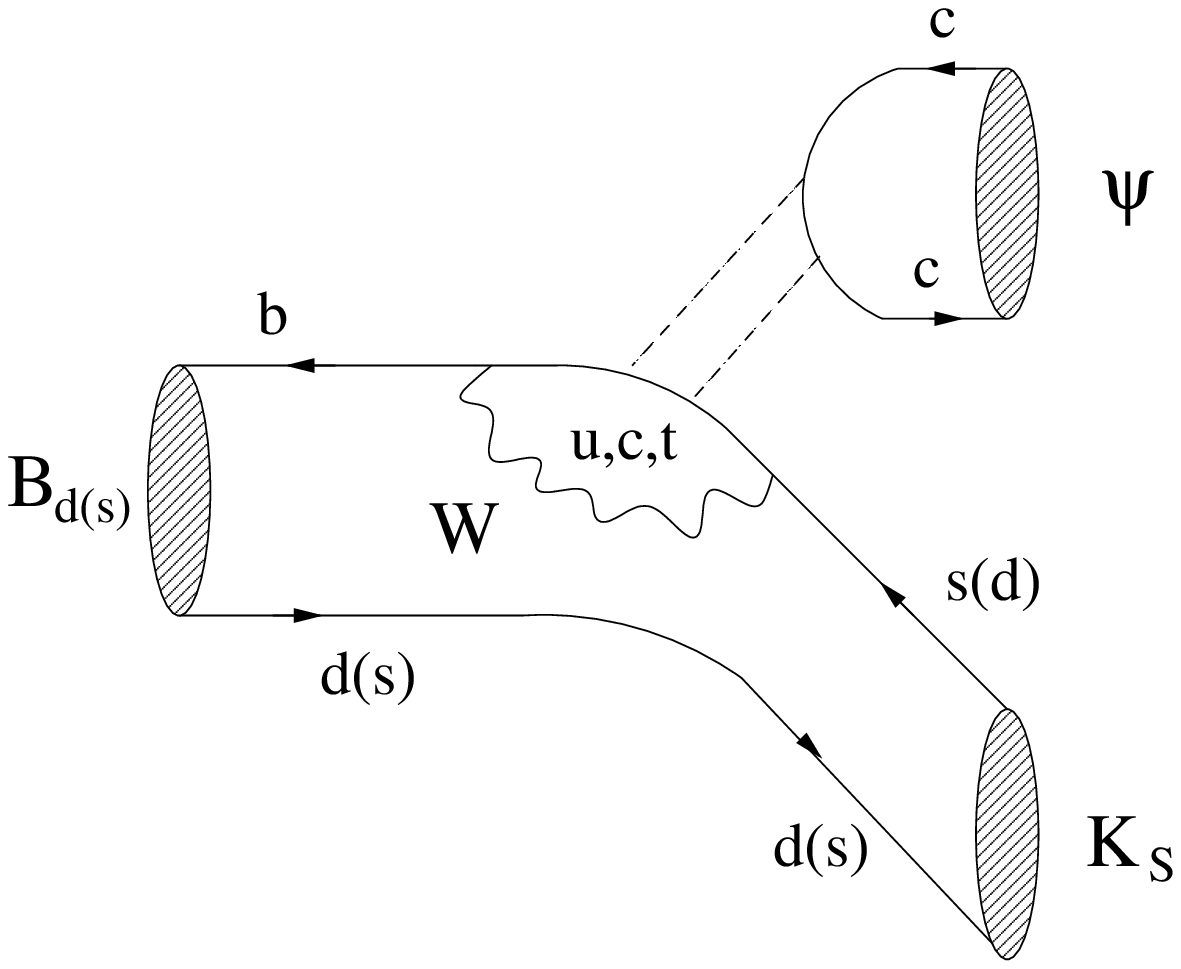}
\end{center}
\caption[]{Feynman diagrams contributing to $B_{d(s)}\to J/\psi\, K_{\rm S}$.
The dashed lines in the penguin topology represent a colour-singlet
exchange.}\label{fig:BsPsiKS}
\end{figure}

As we have already noted in Subsection~\ref{sec:BdPsiKS}, the 
``gold-plated'' mode $B_d\to J/\psi\, K_{\rm S}$ plays an outstanding 
role in the determination of the CP-violating weak 
$B^0_d$--$\overline{B^0_d}$ mixing phase $\phi_d$, i.e.\ of
the CKM angle $\beta$. In this subsection, we will have a closer look at 
$B_s\to J/\psi\, K_{\rm S}$, which is related to $B_d\to J/\psi\, K_{\rm S}$
by interchanging all down and strange quarks, as can be seen
in Fig.~\ref{fig:BsPsiKS}.

Making use of the unitarity of the CKM matrix and applying the Wolfenstein 
parametrization \cite{wolf}, generalized to include non-leading terms in 
$\lambda$ \cite{BLO}, we obtain \cite{BsPsiK}
\begin{equation}\label{Bd-ampl2}
A(B_d^0\to J/\psi\, K_{\rm S})=\left(1-\frac{\lambda^2}{2}\right){\cal A'}
\left[1+\left(\frac{\lambda^2}{1-\lambda^2}\right)a'e^{i\theta'}e^{i\gamma}
\right],
\end{equation}
where
\begin{equation}\label{Aap-def}
{\cal A'}\equiv\lambda^2A\left(A_{\rm cc}^{c'}+A_{\rm pen}^{ct'}\right),
\end{equation}
with $A_{\rm pen}^{ct'}\equiv A_{\rm pen}^{c'}-A_{\rm pen}^{t'}$, and
\begin{equation}\label{ap-def}
a'e^{i\theta'}\equiv R_b\left(1-\frac{\lambda^2}{2}\right)\left(
\frac{A_{\rm pen}^{ut'}}{A_{\rm cc}^{c'}+A_{\rm pen}^{ct'}}\right).
\end{equation}
The amplitudes $A_{\rm cc}^{c'}$ and $A_{\rm pen}^{q'}$ ($q\in\{u,c,t\}$)
describe the current--current, i.e.\ ``tree'', and penguin processes in 
Fig.~\ref{fig:BsPsiKS}, and $A_{\rm pen}^{ut'}$ is defined in analogy to 
$A_{\rm pen}^{ct'}$. On the other hand, the $B_s^0\to J/\psi\, K_{\rm S}$ 
decay amplitude can be parametrized as follows:
\begin{equation}\label{Bs-ampl}
A(B_s^0\to J/\psi\, K_{\rm S})=-\lambda\,{\cal A}\left[1-a\, e^{i\theta}
e^{i\gamma}\right],
\end{equation}
where
\begin{equation}
{\cal A}\equiv\lambda^2A\left(A_{\rm cc}^{c}+A_{\rm pen}^{ct}\right)
\end{equation}
and
\begin{equation}\label{a-def}
a\, e^{i\theta}\equiv R_b\left(1-\frac{\lambda^2}{2}\right)\left(
\frac{A_{\rm pen}^{ut}}{A_{\rm cc}^{c}+A_{\rm pen}^{ct}}\right)
\end{equation}
correspond to (\ref{Aap-def}) and (\ref{ap-def}), respectively. It should
be emphasized that (\ref{Bd-ampl2}) and (\ref{Bs-ampl}) rely only on 
the unitarity of the CKM matrix. In particular, these Standard-Model
parametrizations of the $B_{d(s)}^0\to J/\psi\, K_{\rm S}$ decay 
amplitudes also take into account final-state-interaction effects, which 
can be considered as long-distance penguin topologies with internal up- 
and charm-quark exchanges \cite{bfm}.

If we compare (\ref{Bd-ampl2}) and (\ref{Bs-ampl}) with each other, we
observe that the quantity $a' e^{i\theta'}$ is doubly Cabibbo-suppressed
in the $B_d^0\to J/\psi\, K_{\rm S}$ decay amplitude (\ref{Bd-ampl2}), 
whereas $a\, e^{i\theta}$ enters in the $B_s^0\to J/\psi\, K_{\rm S}$ 
amplitude (\ref{Bs-ampl}) in a Cabibbo-allowed way. Consequently, there
may be sizeable CP-violating effects in $B_s\to J/\psi\, K_{\rm S}$.
As was pointed out in \cite{BsPsiK}, the $U$-spin flavour symmetry of
strong interactions allows us to extract $\gamma$, as well as interesting 
hadronic quantities, from the CP asymmetries ${\cal A}_{\rm CP}^{\rm dir}
(B_s\to J/\psi\, K_{\rm S})$, ${\cal A}_{\rm CP}^{\rm mix}
(B_s\to J/\psi\, K_{\rm S})$ and the CP-averaged 
$B_{d(s)}\to J/\psi\, K_{\rm S}$ branching ratios. The theoretical accuracy 
of this approach is only limited by $U$-spin-breaking corrections, and 
there are no problems due to final-state-interaction effects. As an
interesting by-product, this strategy allows us to take into account the -- 
presumably very small -- penguin contributions in the determination of 
$\phi_d=2\beta$ from $B_d\to J/\psi\, K_{\rm S}$, which is
an important issue in view of the impressive accuracy that can be achieved 
in the LHC era. Moreover, we have an interesting relation between the
direct $B_{s(d)}\to J/\psi\, K_{\rm S}$ CP asymmetries and the corresponding
CP-averaged branching ratios:
\begin{equation}
\frac{{\cal A}_{\rm CP}^{\rm dir}(B_d\to J/\psi\, 
K_{\rm S})}{{\cal A}_{\rm CP}^{\rm dir}(B_s\to J/\psi\, K_{\rm S})}\approx
-\,\frac{\mbox{BR}(B_s\to J/\psi\, K_{\rm S})}{\mbox{BR}(B_d\to J/\psi\, 
K_{\rm S})}\,.
\end{equation}
The experimental feasibility of the extraction of $\gamma$ sketched above
depends strongly on the size of the penguin effects in 
$B_s\to J/\psi\, K_{\rm S}$, which are very hard to estimate. A similar
strategy is provided by $B_{d (s)}\to D^{\,+}_{d(s)}\, D^{\,-}_{d(s)}$ 
decays. For a detailed discussion, the reader is referred to \cite{BsPsiK}.

\boldmath
\subsection{Extracting $\beta$ and $\gamma$ from $B_d\to\pi^+\pi^-$ and\\ 
$B_s\to K^+K^-$}\label{sec:BsKK}
\unboldmath

In this subsection, a new way of making use of the CP-violating 
observables of the decay $B_d\to\pi^+\pi^-$ is discussed \cite{BsKK}: 
combining them with those of $B_s\to K^+K^-$ -- the $U$-spin counterpart of 
$B_d\to\pi^+\pi^-$ -- a simultaneous determination of $\phi_d=2\beta$ and 
$\gamma$ becomes possible. This approach is not affected by any penguin 
topologies -- it rather makes use of them -- and does not rely on 
certain ``plausible'' dynamical or model-dependent assumptions. Moreover, 
final-state-interaction effects, which led to considerable attention in 
the recent literature in the context of the determination of $\gamma$ from 
$B\to\pi K$ decays (see Subsection~\ref{sec:BpiK}), do not lead to any 
problems, and the theoretical accuracy is only limited by $U$-spin-breaking 
effects. This strategy, which is furthermore very promising to search for
indications of new physics \cite{FMat}, is conceptually quite similar to 
the extraction of $\gamma$ from $B_{s(d)}\to J/\psi\, K_{\rm S}$ discussed 
in the previous subsection. However, it appears to be more favourable in view 
of the $U$-spin-breaking effects and the experimental feasibility.

\begin{figure}
\vskip -0.5truein
\begin{center}
\leavevmode
\epsfysize=4.5truecm 
\epsffile{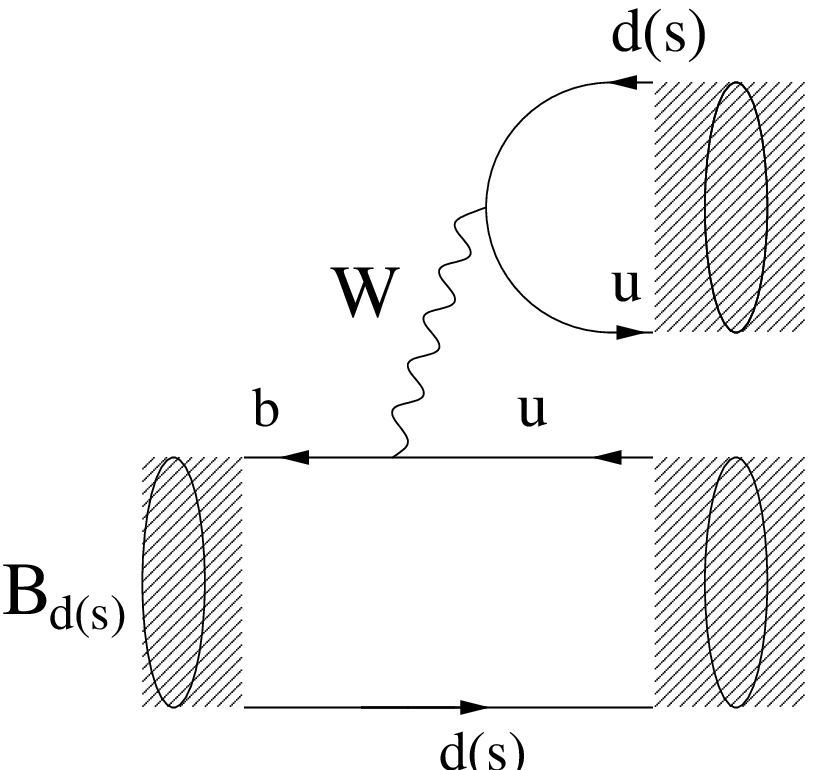} \hspace*{1truecm}
\epsfysize=4.5truecm 
\epsffile{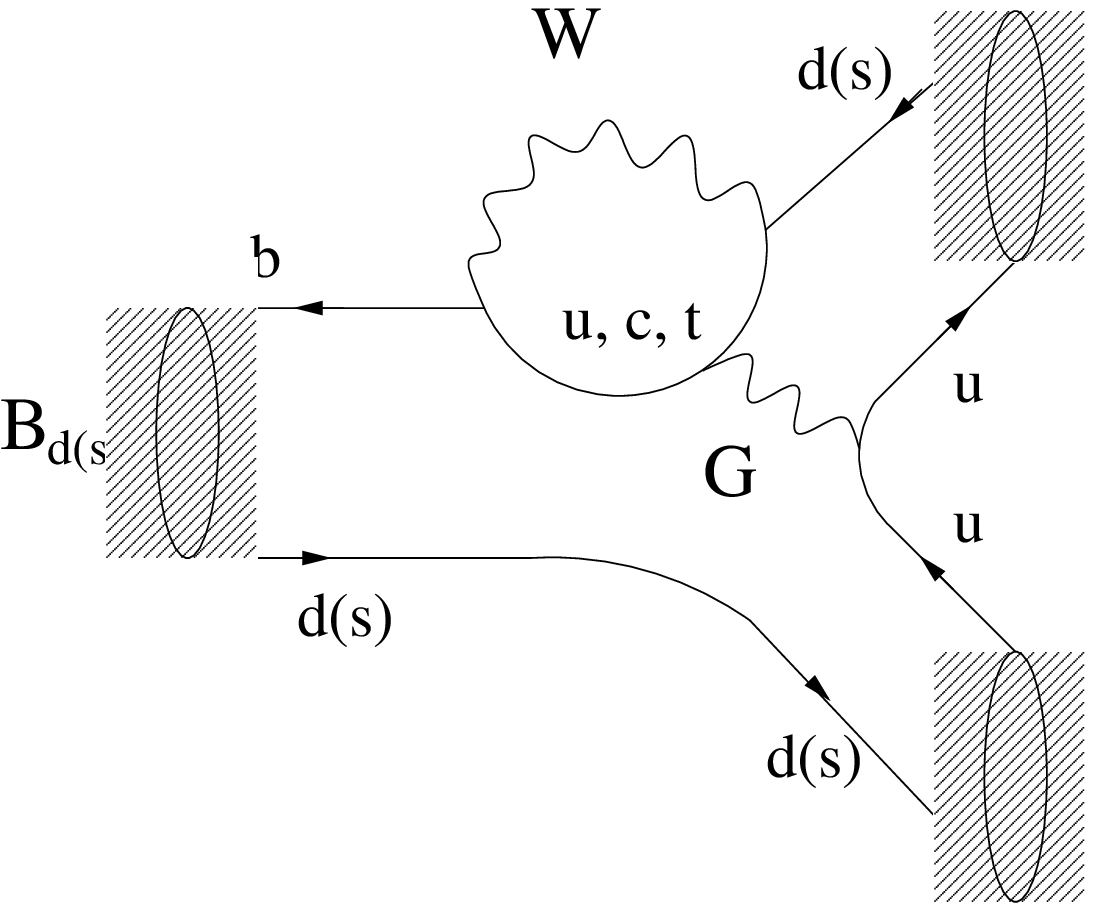}
\end{center}
\caption[]{Feynman diagrams contributing to $B_d\to\pi^+\pi^-$ and
$B_s\to K^+K^-$.}\label{fig:BsKK}
\end{figure}

The leading-order Feynman diagrams contributing to $B_d\to\pi^+\pi^-$ and
$B_s\to K^+K^-$ are shown in Fig.~\ref{fig:BsKK}. If we make use of the 
unitarity of the CKM matrix and apply the Wolfenstein parametrization 
\cite{wolf}, generalized to include non-leading terms in $\lambda$ 
\cite{BLO}, the $B_d^0\to\pi^+\pi^-$ decay amplitude can be expressed 
as follows \cite{BsKK}:
\begin{equation}\label{Bdpipi-ampl}
A(B_d^0\to\pi^+\pi^-)=e^{i\gamma}\left(1-\frac{\lambda^2}{2}\right){\cal C}
\left[1-d\,e^{i\theta}e^{-i\gamma}\right],
\end{equation}
where
\begin{equation}\label{C-def}
{\cal C}\equiv\lambda^3A\,R_b\left(A_{\rm cc}^{u}+A_{\rm pen}^{ut}\right),
\end{equation}
with $A_{\rm pen}^{ut}\equiv A_{\rm pen}^{u}-A_{\rm pen}^{t}$, and
\begin{equation}\label{d-def}
d\,e^{i\theta}\equiv\frac{1}{(1-\lambda^2/2)R_b}
\left(\frac{A_{\rm pen}^{ct}}{A_{\rm cc}^{u}+A_{\rm pen}^{ut}}\right).
\end{equation}
In analogy to (\ref{Bdpipi-ampl}), we obtain for the $B_s^0\to K^+K^-$
decay amplitude 
\begin{equation}\label{BsKK-ampl}
A(B_s^0\to K^+K^-)=e^{i\gamma}\lambda\,{\cal C}'\left[1+\left(
\frac{1-\lambda^2}{\lambda^2}\right)d'e^{i\theta'}e^{-i\gamma}\right],
\end{equation}
where
\begin{equation}
{\cal C}'\equiv\lambda^3A\,R_b\left(A_{\rm cc}^{u'}+A_{\rm pen}^{ut'}\right)
\end{equation}
and 
\begin{equation}\label{dp-def}
d'e^{i\theta'}\equiv\frac{1}{(1-\lambda^2/2)R_b}
\left(\frac{A_{\rm pen}^{ct'}}{A_{\rm cc}^{u'}+A_{\rm pen}^{ut'}}\right)
\end{equation}
correspond to (\ref{C-def}) and (\ref{d-def}), respectively. The general
expressions for the $B_d\to\pi^+\pi^-$ and $B_s\to K^+K^-$ observables
(\ref{ee7}) and (\ref{ADGam}) in terms of the parameters specified above 
can be found in \cite{BsKK}. 

As can be seen in Fig.\ \ref{fig:BsKK}, $B_d\to\pi^+\pi^-$ and
$B_s\to K^+K^-$ are related to each other by interchanging all down and
strange quarks. Consequently, the $U$-spin flavour symmetry of strong
interactions implies
\begin{equation}\label{U-spin-rel}
d'=d\quad\mbox{and}\quad\theta'=\theta.
\end{equation}
If we assume that the $B^0_s$--$\overline{B^0_s}$ mixing phase $\phi_s$ 
is negligibly small, or that it is fixed through $B_s\to J/\psi\,\phi$, 
the four CP-violating observables provided by $B_d\to\pi^+\pi^-$ and 
$B_s\to K^+K^-$ depend -- in the strict $U$-spin limit -- on the four 
``unknowns'' $d$, $\theta$, $\phi_d=2\beta$ and $\gamma$. We have
therefore sufficient observables at our disposal to extract these 
quantities simultaneously. In order to determine $\gamma$, it suffices 
to consider ${\cal A}_{\rm CP}^{\rm mix}(B_s\to K^+K^-)$ and the direct 
CP asymmetries ${\cal A}_{\rm CP}^{\rm dir}(B_s\to K^+K^-)$, 
${\cal A}_{\rm CP}^{\rm dir}(B_d\to\pi^+\pi^-)$. If we make use, in addition,
of ${\cal A}_{\rm CP}^{\rm mix}(B_d\to\pi^+\pi^-)$, $\phi_d$ can be determined 
as well. The formulae to implement this approach in a mathematical way
are given in \cite{BsKK}.

\begin{figure}[t]
\centerline{\rotate[r]{
\epsfysize=10truecm
{\epsffile{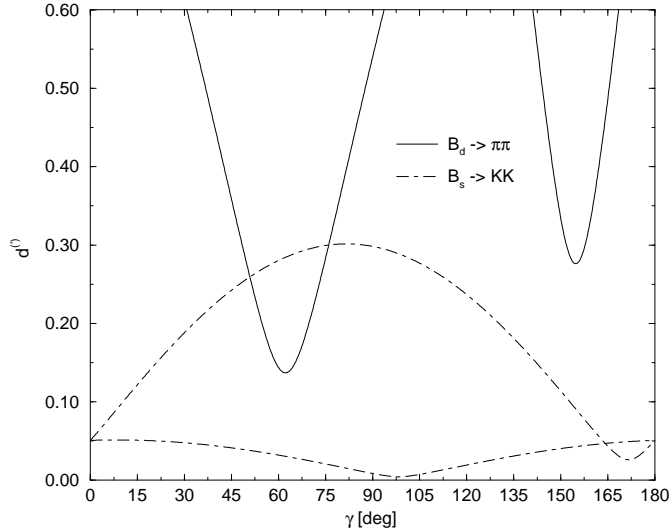}}}}
\caption{The contours in the $\gamma$--$d^{(')}$ planes fixed through the
$B_d\to\pi^+\pi^-$, $B_s\to K^+K^-$ observables for a specific example 
discussed in the text.}\label{fig:BsKKcont}
\end{figure}

If we use the $B^0_d$--$\overline{B^0_d}$ mixing phase as an input, 
there is a different way of combining ${\cal A}_{\rm CP}^{\rm dir}
(B_d\to\pi^+\pi^-)$, ${\cal A}_{\rm CP}^{\rm mix}(B_d\to\pi^+\pi^-)$ with 
${\cal A}_{\rm CP}^{\rm dir}(B_s\to K^+K^-)$, 
${\cal A}_{\rm CP}^{\rm mix}(B_s\to K^+K^-)$. The point is that these
observables allow us to fix contours in the $\gamma$--$d$ and 
$\gamma$--$d'$ planes as functions of the $B^0_d$--$\overline{B^0_d}$ and 
$B^0_s$--$\overline{B^0_s}$ mixing phases in a {\it theoretically clean} 
way. In Fig.~\ref{fig:BsKKcont}, these contours are shown for a specific
example \cite{BsKK}:
\begin{equation}\label{obs-examp}
\begin{array}{lcllcl}
{\cal A}_{\rm CP}^{\rm dir}(B_d\to\pi^+\pi^-)&=&+24\%,\,\,& 
{\cal A}_{\rm CP}^{\rm mix}(B_d\to\pi^+\pi^-)&=&+4.4\%,\\ 
{\cal A}_{\rm CP}^{\rm dir}(B_s\to K^+K^-)&=&-17\%,\,\,&
{\cal A}_{\rm CP}^{\rm mix}(B_s\to K^+K^-)&=&-28\%,
\end{array}
\end{equation}
corresponding to the input parameters $d=d'=0.3$, $\theta=\theta'=210^\circ$,
$\phi_s=0$, $\phi_d=53^\circ$ and $\gamma=76^\circ$. In order to extract 
$\gamma$ and the hadronic parameters $d$, $\theta$, $\theta'$ with the help 
of these contours, the $U$-spin relation $d'=d$ is sufficient. The 
intersection of the contours shown in Fig.~\ref{fig:BsKKcont} yields a 
twofold solution for $\gamma$, given by $51^\circ$ and our input value 
of $76^\circ$. The resolution of this ambiguity is discussed in \cite{BsKK}. 
A first experimental feasibility study for LHCb, using the set of observables
given in (\ref{obs-examp}), gave an uncertainty of 
$\left.\Delta\gamma\right|_{\rm exp}=2.3^\circ$ for five years of 
data taking and looks very promising \cite{wilkinson}.

It should be emphasized that the theoretical accuracy of $\gamma$ and
of the hadronic parameters $d$, $\theta$ and $\theta'$ is only limited
by $U$-spin-breaking effects. In particular, it is not affected by
any final-state-interaction or penguin effects. A first consistency check
is provided by $\theta=\theta'$. Moreover, we may determine the normalization
factors ${\cal C}$ and ${\cal C}'$ of the $B^0_d\to\pi^+\pi^-$ and
$B^0_s\to K^+K^-$ decay amplitudes (see (\ref{Bdpipi-ampl}) and 
(\ref{BsKK-ampl})) with the help of the corresponding CP-averaged
branching ratios. Comparing them with the ``factorized'' result
\begin{equation}
\left|\frac{{\cal C}'}{{\cal C}}\right|_{\rm fact}=\,
\frac{f_K}{f_\pi}\frac{F_{B_sK}(M_K^2;0^+)}{F_{B_d\pi}(M_\pi^2;0^+)}
\left(\frac{M_{B_s}^2-M_K^2}{M_{B_d}^2-M_\pi^2}\right),
\end{equation}
we have another interesting probe for $U$-spin-breaking effects. 
Interestingly, $d'e^{i\theta'}=d\,e^{i\theta}$ is not affected by 
$U$-spin-breaking corrections within a certain model-dependent approach 
(a modernized version of the ``Bander--Silverman--Soni mechanism'' 
\cite{bss}), making use -- among other things -- of the ``factorization'' 
hypothesis to estimate the relevant hadronic matrix elements \cite{BsKK}. 
Although this approach seems to be rather simplified and may be affected 
by non-factorizable effects, it strengthens our confidence into the $U$-spin 
relations used for the extraction of $\beta$ and $\gamma$ from the decays 
$B_d\to\pi^+\pi^-$ and $B_s\to K^+K^-$.

The strategy discussed in this subsection is very promising for 
second-generation $B$-physics experiments at hadron machines, 
where the physics potential of the $B_s$ system can be fully exploited. 
At the asymmetric $e^+e^-$ $B$-factories operating at the $\Upsilon(4S)$ 
resonance, BaBar and BELLE, which have already seen the first events, 
this is unfortunately not possible. However, there is also a variant of 
the strategy to determine $\gamma$, where $B_d\to\pi^\mp K^\pm$ is used 
instead of $B_s\to K^+K^-$ \cite{BsKK}. This approach has the advantage 
that all required time-dependent measurements can in principle be performed 
at the asymmetric $e^+e^-$ machines. On the other hand, it relies -- in 
addition to the $SU(3)$ flavour symmetry -- on the smallness of certain 
``exchange'' and ``penguin annihilation'' topologies, which may be enhanced 
by final-state-interaction effects. Consequently, its theoretical accuracy 
cannot compete with the ``second-generation'' $B_d\to\pi^+\pi^-$, 
$B_s\to K^+K^-$ approach, which is not affected by such problems.

\boldmath
\subsection{Extracting CKM Phases and Hadronic Parameters from
Angular Distributions of $B_{d,s}$ Decays}\label{sec:ang}
\unboldmath

A very interesting laboratory to explore CP violation and the
hadronization dynamics of non-leptonic $B$ decays is provided 
by quasi-two-body modes $B_q\to X_1\,X_2$ of neutral $B_q$-mesons, 
where both $X_1$ and $X_2$ carry spin and continue to decay through 
CP-conserving interactions \cite{FD,ang-stud}. In this case, the 
time-dependent angular distribution of the decay products of $X_1$ 
and $X_2$ provides valuable information. For an initially, i.e.\ at 
time $t=0$, present $B_q^0$-meson, it can be written as 
\begin{equation}\label{ang}
f(\Theta,\Phi,\Psi;t)=\sum_k{\cal O}^{(k)}(t)g^{(k)}(\Theta,\Phi,\Psi),
\end{equation}
where we have denoted the angles describing the kinematics of the decay
products of $X_1$ and $X_2$ generically by $\Theta$, $\Phi$ and
$\Psi$. There are two different kinds of observables ${\cal O}^{(k)}(t)$,
describing the time evolution of the angular distribution (\ref{ang}):
observables $\left|A_f(t)\right|^2$, corresponding to ``ordinary'' decay
rates, and interference terms of the type
\begin{equation}\label{inter}
\mbox{Re}[A_{\tilde f}^\ast(t)A_f(t)],\quad \mbox{Im}[A_{\tilde f}^\ast(t)
A_f(t)],
\end{equation}
where the amplitudes $A_f(t)$ correspond to a given final-state configuration 
$[X_1\,X_2]_f$. In comparison with strategies using $B_q\to P_1\,P_2$ 
decays into two pseudoscalar mesons, the angular distributions of the 
$B_q\to X_1\,X_2$ modes provide many more cross-checks and allow, in 
certain cases, the resolution of discrete ambiguities, which usually
affect the extraction of CKM phases. The latter feature is due to the 
observables (\ref{inter}). 

In a recent paper \cite{RF-ang}, I presented the general formalism to 
extract CKM phases and hadronic parameters from the time-dependent angular 
distributions (\ref{ang}) of certain $B_q\to X_1\,X_2$ decays, taking 
also into account penguin contributions. If we fix the mixing phase $\phi_q$ 
separately, it is possible to determine a CP-violating weak phase $\omega$, 
which is usually given by the angles of the unitarity triangle shown in 
Fig.~\ref{fig:UT}\,(a), and interesting hadronic quantities as a function 
of a {\it single} hadronic parameter (this feature is also discussed in 
another recent paper \cite{LSS}). If we determine this parameter, for 
instance, by comparing $B_q\to X_1\,X_2$ with an $SU(3)$-related mode, 
all remaining parameters, including $\omega$, can 
be extracted. If we are willing to make more extensive use of 
flavour-symmetry arguments, it is in principle possible to determine 
the $B^0_q$--$\overline{B^0_q}$ mixing phase $\phi_q$ as well. As the 
technical details of this approach are rather involved, let us just have 
a brief look at some of its applications.

\boldmath
\subsubsection{$B_d\to J/\psi\,\rho^0$ and $B_s\to J/\psi\, \phi$}
\unboldmath

The structure of the decay amplitudes of these modes is very similar
to the ones of $B_s\to J/\psi\,K_{\rm S}$ and $B_d\to J/\psi\,K_{\rm S}$ 
discussed in Subsection~\ref{sec:BsPsiKS}. They can be related to 
each other through $SU(3)$ and certain dynamical arguments, involving 
``exchange'' and ``penguin annihilation'' topologies, and allow the extraction
of the $B^0_d$--$\overline{B^0_d}$ mixing phase $\phi_d=2\beta$. Because
of the interference effects leading to the observables (\ref{inter}), 
both $\sin\phi_d$ and $\cos\phi_d$ can be determined, thereby allowing us to 
fix $\phi_d$ {\it unambiguously}. As we have seen above, this phase is an
important input for several strategies to determine $\gamma$. For alternative
methods to resolve the twofold ambiguity arising in the extraction of 
$\phi_d$ from ${\cal A}^{\mbox{{\scriptsize mix--ind}}}_{\mbox{{\scriptsize
CP}}}(B_d\to J/\psi\, K_{\mbox{{\scriptsize S}}})=-\sin\phi_d$, the reader 
is referred to \cite{ambig}.

Should the penguin effects in $B_d\to J/\psi\,\rho^0$ be sizeable, 
$\gamma$ can be determined as well. As an interesting by-product, this 
strategy allows us to take into account the penguin effects in the 
extraction of the $B^0_s$--$\overline{B^0_s}$ mixing phase from 
$B_s\to J/\psi\,\phi$, which is an important issue for the LHC era. Moreover, 
valuable insights into $SU(3)$-breaking effects can be obtained.

\boldmath
\subsubsection{$B_d\to\rho^+\rho^-$ and 
$B_s\to K^{\ast+}\,K^{\ast-}$}
\unboldmath

The structure of the decay amplitudes of these transitions is completely 
analogous to the ones of $B_d\to\pi^+\pi^-$ and $B_s\to K^+K^-$ discussed
in Subsection~\ref{sec:BsKK}. They can be related to each other through 
$U$-spin arguments, thereby allowing the extraction of $\gamma$ and 
of the $B^0_d$--$\overline{B^0_d}$ and $B^0_s$--$\overline{B^0_s}$
mixing phases. In contrast to the $B_d\to\pi^+\pi^-$, $B_s\to K^+K^-$
strategy, both mixing phases can in principle be determined, and
many more cross-checks of interesting $U$-spin relations can be performed.

\boldmath
\subsubsection{$B_d\to K^{\ast0}\,\overline{K^{\ast0}}$ and 
$B_s\to K^{\ast0}\,\overline{K^{\ast0}}$}
\unboldmath

These decays are also $U$-spin counterparts and allow the simultaneous
extraction of $\gamma$, $\phi_d$ and $\phi_s$. As they are pure
penguin-induced modes, they are very sensitive to new physics. A 
particular parametrization of the $B_d\to K^{\ast0}\,\overline{K^{\ast0}}$ 
decay amplitude allows us to probe also the weak phase 
$\phi\equiv\phi_d-2\beta$. Within the Standard Model, we have 
$\phi=0$. However, this relation may well be affected by new physics, 
and represents an interesting test of the Standard-Model 
description of CP violation. Therefore it would be very important to 
determine this combination of CKM phases experimentally. The observables 
of the $B_d\to K^{\ast0}[\to\pi^-K^+]\,\overline{K^{\ast0}}[\to\pi^+K^-]$ 
angular distribution may provide an important step towards this goal.

\vspace*{0.5truecm}

\noindent
Since the formalism presented in \cite{RF-ang}, which we have sketched
in this subsection, is very general, it can be applied to many other 
decays. Detailed studies are required to explore which channels are most 
promising from an experimental point of view.

\newpage

\section{Conclusions and Outlook}\label{sec:concl}

In conclusion, we have seen that the phenomenology of non-leptonic decays 
of $B$-mesons is very rich and provides a fertile testing ground for the 
Standard-Model description of CP violation. Research has been very
active in this field over the last couple of years, and we have 
discussed some of the most recent theoretical developments, including  
determinations of $\gamma$ from $B\to\pi K$ and $B_{s(d)}\to J/\psi\, 
K_{\rm S}$ decays, an extraction of $\beta$ and $\gamma$, which is offered 
by $B_d\to \pi^+\pi^-$ and $B_s\to K^+K^-$, and a general approach to extract 
CKM phases and hadronic parameters from angular distributions of certain 
non-leptonic decays of $B_{d,s}$-mesons. In these new strategies, a strong 
emphasis was given to the $B_s$ system, which has a very powerful physics 
potential and is of particular interest for $B$-physics experiments at 
hadron machines. 

The $B$-factory era in particle physics has just started, as the BaBar and
BELLE detectors have recently observed their first events. In the near future,
CLEO-III, HERA-B and CDF-II will also start taking data, and the first
results will certainly be very exciting. However, in order to establish
the presence of physics beyond the Standard Model, it may well be that we
have to wait for second-generation $B$-physics experiments at hadron 
machines such as LHCb or BTeV, which are expected to start operation 
around 2005. Hopefully, these experiments will bring several unexpected 
results, leading to an exciting and fruitful interaction between theorists 
and experimentalists!

\vspace*{0.5truecm}

\noindent
{\bf{\it Acknowledgements}}

\vspace*{0.2truecm}

\noindent
I would like to thank the organizers for inviting me to that stimulating
conference in a very enjoyable environment.


\begin{thebibliography}{99}

\bibitem{ckm}N. Cabibbo, {\it Phys.\ Rev.\ Lett.}~{\bf 10} (1963) 531; 
M. Kobayashi and T.~Maskawa, {\it Progr.\ Theor.\ Phys.}~{\bf 49} (1973) 652.

\bibitem{AKL}R. Aleksan, B. Kayser and D. London, {\it Phys.\ Rev.\ 
Lett.}~{\bf 73} (1994) 18.

\bibitem{wolf}L. Wolfenstein, {\it Phys.\ Rev.\ Lett.}~{\bf 51} (1983) 1945.

\bibitem{ut}L.L. Chau and W.-Y. Keung, {\it Phys.\ Rev.\ Lett.}~{\bf 53} 
(1984) 1802; C. Jarlskog and R.~Stora, {\it Phys.\ Lett.}~{\bf B208} (1988) 
268.

\bibitem{BLO}A.J. Buras, M.E. Lautenbacher and G. Ostermaier, {\it Phys.\ 
Rev.}~{\bf D50} (1994) 3433.

\bibitem{rev}For a review, see R. Fleischer, {\it Int.\ J. Mod.\ 
Phys.}~{\bf A12} (1997) 2459.

\bibitem{dun}I. Dunietz, {\it Phys.\ Rev.}~{\bf D52} (1995) 3048.

\bibitem{DGamma-cal}For a recent calculation of $\Delta\Gamma_s$, see 
M. Beneke, G. Buchalla, C.~Greub, A. Lenz and U. Nierste, {\it Phys.\ 
Lett.}~{\bf B459} (1999) 631.

\bibitem{bisa}A.B. Carter and A.I. Sanda, {\it Phys.\ Rev.\ Lett.}~{\bf 45}
(1980) 952; {\it Phys.\ Rev.}~{\bf D23} (1981) 1567; I.I. Bigi and A.I. 
Sanda, {\it Nucl.\ Phys.}~{\bf B193} (1981) 85.

\bibitem{nir-sil}Y. Nir and D. Silverman, {\it Nucl.\ Phys.}~{\bf B345}
(1990) 301.

\bibitem{sin2b-exp}OPAL Collaboration (K. Ackerstaff {\it et al.}), {\it Eur.\ 
Phys.\ J.}~{\bf C5} (1998) 379; CDF Collaboration (F. Abe {\it et al.}),
{\it Phys.\ Rev.\ Lett.}~{\bf 81} (1998) 5513; for an updated analysis,
see preprint CDF/PUB/BOTTOM/CDF/4855, and the contribution by Petar Maksimovic
to these proceedings.

\bibitem{BsPsiK}R. Fleischer, {\it Eur.\ Phys.\ J.}~{\bf C} (1999) 
DOI 10.1007/s100529900099 [hep-ph/9903455].

\bibitem{alpha-uncert}See, for instance, M. Gronau, 
{\it Phys.\ Lett.}~{\bf B300} (1993) 163; J.P. Silva and L. Wolfenstein, 
{\it Phys.\ Rev.}~{\bf D49} (1994) R1151; R. Aleksan {\it et al.}, 
{\it Phys.\ Lett.}~{\bf B356} (1995) 95; A.J. Buras and R. Fleischer, 
{\it Phys.\ Lett.}~{\bf B360} (1995) 138; F. DeJongh and P. Sphicas, 
{\it Phys.\ Rev.}~{\bf D53} (1996) 4930; M. Ciuchini {\it et al.}, 
{\it Nucl.\ Phys.}~{\bf B501} (1997) 271; P.S.~Marrocchesi and N. Paver, 
{\it Int.\ J. Mod.\ Phys.}~{\bf A13} (1998) 251; A. Ali, G.~Kramer and 
C.-D. L\"u, {\it Phys.\ Rev.}~{\bf D59} (1999) 014005.

\bibitem{charles}J. Charles, {\it Phys.\ Rev.}~{\bf D59} (1999) 054007.

\bibitem{gl}M. Gronau and D. London, {\it Phys.\ Rev.\ Lett.}~{\bf 65} (1990)
3381.

\bibitem{BF}A.J. Buras and R. Fleischer, preprint CERN-TH/98-319 (1998)
[hep-ph/9810260], to appear in {\it Eur.\ Phys.\ J.}~{\bf C}.

\bibitem{GPY}M. Gronau, D. Pirjol and T.-M. Yan, preprint CLNS-98-1582 (1998)
[hep-ph/9810482]. 

\bibitem{gq-alpha}Y. Grossman and H.R. Quinn, {\it Phys.\ Rev.}~{\bf D58},
017504 (1998). 

\bibitem{Brhopi}H. Lipkin, Y. Nir, H. Quinn and A. Snyder, {\it Phys.\
Rev.}~{\bf D44} (1991)  1454; A. Snyder and H. Quinn, {\it Phys.\ 
Rev.}~{\bf D48} (1993) 2139.

\bibitem{BsKK}R. Fleischer, {\it Phys.\ Lett.}~{\bf B459} (1999) 306.

\bibitem{cleo-Bpipi}D.E. Jaffe (CLEO Collaboration), talk given at the
{\it 8th International Symposium on Heavy Flavour Physics}, Southampton, 
25--29 July 1999.

\bibitem{BBNS}M. Beneke, G. Buchalla, M. Neubert and C.T. Sachrajda,
preprint CERN-TH/99-126 (1999) [hep-ph/9905312].

\bibitem{BDpi}R.G. Sachs, preprint EFI-85-22 (1985) (unpublished);
I. Dunietz and R.G. Sachs, {\it Phys.\ Rev.}~{\bf D37} (1988) 3186 [E:
{\it Phys.\ Rev.}~{\bf D39} (1989) 3515]; I. Dunietz, {\it Phys.\ 
Lett.}~{\bf B427} (1998) 179.

\bibitem{BDpi-exp}Experimental feasibility studies were performed by
J. Gronberg and H.~Nelson for the {\it The BaBar Phyics Book}, eds.\
P.F. Harison and H.R.~Quinn (SLAC report 504, October 1998), and by
J. Rademacker and G.~Wilkinson for the Workshop on {\it Standard Model 
Physics (and more) at the LHC}, CERN (1999).

\bibitem{BF-rev}A.J. Buras and R. Fleischer, in {\it Heavy Flavours II},
eds.\ A.J. Buras and M. Lindner (World Scientific, Singapore, 1998)
[hep-ph/9704376].

\bibitem{FD}R. Fleischer and I. Dunietz, 
{\it Phys.\ Lett.}~{\bf B387} (1996) 361 
and {\it Phys.\ Rev.}~{\bf D55} (1997) 259.

\bibitem{adk}R. Aleksan, I. Dunietz and B. Kayser, 
{\it Z. Phys.}~{\bf C54} (1992) 653.

\bibitem{ddf1}A.S. Dighe, I. Dunietz and R. Fleischer, {\it Eur.\ 
Phys.\ J}~{\bf C6} (1999) 647.

\bibitem{RF-ang}R. Fleischer, preprint CERN-TH/99-92 (1999) 
[hep-ph/9903540v2], to appear in {\it Phys.\ Rev.}~{\bf D}.

\bibitem{bbmr}G. Barenboim, J. Bernabeu, J. Matias and M. Raidal,  
{\it Phys.\ Rev.}~{\bf D60} (1999) 016003.

\bibitem{smizanska}M. Smizanska, these proceedings.

\bibitem{calvetti}M. Calvetti, these proceedings.

\bibitem{gw}M. Gronau and D. Wyler, {\it Phys.\ Lett.}~{\bf B265} (1991) 172;
see also I.~Dunietz, {\it Phys.\ Lett.}~{\bf B270} (1991) 75.

\bibitem{GRL}M. Gronau, J.L. Rosner and D. London, {\it Phys.\ Rev.\
Lett.}~{\bf 73} (1994) 21.

\bibitem{GHLR}O.F. Hern\'andez, D. London, M. Gronau  and  J.L. Rosner,
{\it Phys.\ Lett.}~{\bf B333} (1994) 500; {\it Phys.\ Rev.}~{\bf D50}
(1994) 4529.

\bibitem{ads}D. Atwood, I. Dunietz and A. Soni, {\it Phys.\ Rev.\ 
Lett.}~{\bf 78} (1997) 3257.

\bibitem{cleo-bdk}CLEO Collaboration (M. Athanas {\it et al.}), {\it Phys.\
Rev.\ Lett.}~{\bf 80} (1998) 5493.

\bibitem{berkelman}K. Berkelman, these proceedings.

\bibitem{FM}R. Fleischer and T. Mannel, {\it Phys.\ Rev.}~{\bf D57} (1998) 
2752.

\bibitem{PAPIII}R. Fleischer, {\it Phys.\ Lett.}~{\bf B365} (1996) 399.

\bibitem{GroRo}M. Gronau and J.L. Rosner, {\it Phys.\ Rev.}~{\bf D57} 
(1998) 6843.

\bibitem{NR}M. Neubert and J.L. Rosner, {\it Phys.\ Lett.}~{\bf B441} (1998)  
403 and {\it Phys.\ Rev.\ Lett.}~{\bf 81} (1998) 5076; M. Neubert,
{\it JHEP} 9902: 014, 1999.

\bibitem{defan}R. Fleischer, {\it Eur.\ Phys.\ J.}~{\bf C6} (1999) 451.

\bibitem{bfm}A.J. Buras, R. Fleischer and T. Mannel, {\it Nucl.\ 
Phys.}~{\bf B533} (1998) 3.

\bibitem{FSI}L. Wolfenstein, {\it Phys.\ Rev.}~{\bf D52} (1995) 537;
J.-M. G\'erard and J. Weyers, {\it Eur.\ Phys.\ J.}~{\bf C7} (1999) 1;
A.F. Falk {\it et al.}, {\it Phys.\ Rev.}~{\bf D57} (1998) 4290;
D. Atwood and A. Soni, {\it Phys.\ Rev.}~{\bf D58} (1998) 036005.

\bibitem{neubert}M. Neubert, {\it Phys.\ Lett.}~{\bf B424} (1998) 152.

\bibitem{BKK}R. Fleischer, {\it Phys.\ Lett.}~{\bf B435} (1998) 221.

\bibitem{groro-FSI}M. Gronau and J. Rosner, {\it Phys.\ Rev.}~{\bf D58} (1998)
113005.

\bibitem{pirjol}D. Pirjol, these proceedings.

\bibitem{dh}N.G. Deshpande and X.-G. He, {\it Phys.\ Rev.\ 
Lett.}~{\bf 74} (1995) 26.

\bibitem{rf-ewp}R. Fleischer, {\it Z. Phys.}~{\bf C62} (1994) 81;
{\it Phys.\ Lett.}~{\bf B321} (1994) 259.

\bibitem{FSI-recent}M. Gronau and D. Pirjol, {\it Phys.\ Lett.}~{\bf B449}
(1999) 321 and preprint CLNS-99-1604 (1999) [hep-ph/9902482];
K. Agashe and N.G. Deshpande, {\it Phys.\ Lett.}~{\bf B451} (1999) 215 and 
{\bf B454} (1999) 359.

\bibitem{new-phys}For reviews, see Y. Grossman, Y. Nir and R. Rattazzi, 
in {\it Heavy Flavours II}, eds.\ A.J. Buras and M. Lindner (World Scientific, 
Singapore, 1998) [hep-ph/9701231]; M. Gronau and D. London, 
{\it Phys.\ Rev.}~{\bf D55} (1997) 2845; Y. Nir and H.R. Quinn, 
{\it Annu.\ Rev.\ Nucl.\ Part.\ Sci.}~{\bf 42} (1992) 211; L. Wolfenstein, 
{\it Phys.\ Rev.}~{\bf D57} (1998) 6857.

\bibitem{FMat}R. Fleischer and J. Matias, preprint CERN-TH/99-164 (1999)
[hep-ph/9906274].

\bibitem{wilkinson}G. Wilkinson, LHCb study for the Workshop on {\it Standard 
Model Physics (and more) at the LHC}, CERN (1999).

\bibitem{bss}M. Bander, D. Silverman and A. Soni, {\it Phys.\ Rev.\
Lett.}~{\bf 43} (1979) 242.

\bibitem{ang-stud}I. Dunietz, H. Quinn, A. Snyder, W. Toki and H.J. Lipkin,
{\it Phys.\ Rev.}~{\bf D43} (1991) 2193.

\bibitem{LSS}D. London, N. Sinha and R. Sinha, preprint UDEM-GPP-TH-99-61 
(1999) [hep-ph/9905404].

\bibitem{ambig}See, for example, Y. Grossman and H.R. Quinn, 
{\it Phys.\ Rev.}~{\bf D56} (1997) 7259; J.~Charles, A. Le Yaouanc, 
L. Oliver, O. P\`ene and J.-C. Raynal, {\it Phys.\ Lett.}~{\bf B425} (1998) 
375 and {\it Phys.\ Rev.}~{\bf D58} (1998) 114021; A.S. Dighe, I. Dunietz and 
R.~\mbox{Fleischer}, {\it Phys.\ Lett.}~{\bf B433} (1998) 147.




\end{thebibliography}
\end{document}